\documentclass[twocolumn, trackchanges]{aastex631}
\usepackage{amsmath}
\usepackage{float}
\usepackage[caption = false]{subfig}
\usepackage{graphicx}
\usepackage{hyperref}

\begin{document}

\title{Neutron-capture Element Abundances of 491 Stars in Milky Way Dwarf Satellite Galaxies from Medium-Resolution Spectra}

\author[0000-0002-0786-7307]{Lauren E. Henderson}
\affiliation{Department of Physics and Astronomy, University of Notre Dame,
225 Nieuwland Science Hall,
Notre Dame, IN 46556, USA}

\author[0000-0001-6196-5162]{Evan N. Kirby}
\affiliation{Department of Physics and Astronomy, University of Notre Dame,
225 Nieuwland Science Hall,
Notre Dame, IN 46556, USA}

\author[0000-0002-4739-046X]{Mithi A. C. de los Reyes}
\affiliation{Department of Physics and Astronomy, Amherst College,
6 East Drive,
Amherst, MA 01002, USA}

\author[0000-0003-0398-639X]{Roman Gerasimov}
\affiliation{Department of Physics and Astronomy, University of Notre Dame,
225 Nieuwland Science Hall,
Notre Dame, IN 46556, USA}

\author[0000-0002-7113-0262]{Viraj Manwadkar}
\affiliation{Department of Physics, Stanford University, 382 Via Pueblo Mall, Stanford, CA 94305, USA}
\affiliation{Kavli Institute for Particle Astrophysics and Cosmology, SLAC National Accelerator Laboratory, 2575 Sand Hill Road, Menlo Park, CA 94025, USA}



\begin{abstract}

The chemical compositions of evolved stars in Local Group dwarf spheroidal galaxies (dSphs) provide insight into the galaxy’s past star formation and nucleosynthesis. Neutron-capture element abundances are especially interesting. In particular, $s$-process elements can provide a third chemical clock for resolving star formation histories in addition to core collapse and Type Ia supernovae.  Likewise, the primary sites of the $r$-process are still areas of extensive research. Until now, the number of stars with neutron-capture element abundances in dSphs has been limited by the need for stars bright enough for high-resolution spectroscopy. We present abundance measurements of the neutron-capture elements Sr, Y, Ba, and Eu with errors $<$ 0.4 dex---as well as new measurements of Mg---in 491 stars in Sculptor, Fornax, Draco, Sextans, and Ursa Minor. The large number of stars in our sample is possible because we used medium-resolution spectra from the DEIMOS spectrograph, assembling the largest homogeneous set of neutron-capture abundances in dwarf spheroidal galaxies to date. By utilizing the abundances of both $s$- and $r$-process elements, we find evidence of an $s$-process contribution at early times in Sculptor from our measurements of [Ba/Fe]. This is a potential signature of $s$-process nucleosynthesis in fast-rotating massive stars. By comparing our measurements of [Eu/Fe] with [Mg/Fe], we show the need for an $r$-process source that has a short delay time to enrich stars in the dSphs. Thus, neutron star mergers are likely not the sole source of $r$-process material in dSphs.

\end{abstract}



\section{Introduction} \label{sec:intro}

 Dwarf galaxies in the Local Group are excellent laboratories for studying galactic evolution. Understanding these stellar populations can yield insight into topics that impact all galaxies, ranging from constraints on dark matter models \citep[e.g.,][]{boylan-kolchin_milky_2012,the_fermi-lat_collaboration_searching_2015}, to analogs to the first galaxies at high redshift \citep[e.g.,][]{cooke_most_2015, flores_local_2021}, to the origins of the periodic table of the elements \citep[e.g.,][]{shetrone_abundance_2001, geisler_sculptor-ing_2005, ji_r-process_2016}.

Dwarf spheroidal galaxies (dSphs) are especially suited for studying galactic chemical evolution at early times; dSphs are relatively old and have abundance patterns with few features (unlike larger galaxies like the Milky Way). The old stellar populations in dSphs contain a record of how and when elements are made and when these galaxies were undergoing star formation.



Chemical evolution is often inferred from the elements released in core-collapse supernovae (CCSNe), especially the $\alpha$ elements (O, Ne, Mg, Si, S, Ar, Ca, and Ti), and those released in Type Ia supernovae, known as the iron-peak elements (V, Cr, Mn, Fe, Co, and Ni). For example, [$\alpha$/Fe] has been used to estimate the star formation histories (SFHs) of dSphs \citep[e.g.,][]{tolstoy_vltuves_2003, kirby_multi-element_2011-1, deboer_fornax_2012, hasselquist_apogee_2021, delosreyes_simultaneous_2022}. Accurate SFHs of nearby dSphs can provide insight into the evolution of the Local Group and the impact of reionization on the star formation of small galaxies \citep{brown_quenching_2014, ricotti_formation_2005}.

Elements heavier than Zn, which are mostly formed via neutron-capture nucleosynthesis, provide additional information because they are often formed on different timescales and in different conditions than lighter elements. There are two main types of neutron-capture nucleosynthesis: the rapid ($r$)-process and the slow ($s$)-process.

In the $r$-process, neutron capture occurs too quickly for the nuclei to $\beta$-decay between captures. This process forms heavy, neutron-rich nuclei and can occur only in extreme environments that are highly energetic and have a large neutron flux. Although neutron star mergers are confirmed sites of the $r$-process thanks to the combined gravitational wave and corresponding multiwavelength observations of the neutron star merger GW170817 \citep{abbott_gw170817_2017, coulter_swope_2017, drout_light_2017, kasen_origin_2017, kilpatrick_electromagnetic_2017, smartt_kilonova_2017, tanaka_kilonova_2017}, there is still ongoing research as to whether they are the sole or most common way to produce $r$-process elements \citep[e.g.,][]{ji_lanthanide_2019, cote_neutron_2019, ezzeddine_rprocess_2020, kobayashi_origin_2020, skuladottir_evidence_2020, vandevoort_neutron_2020, ou_rise_2024}. This ambiguity makes $r$-process elements problematic for estimating star formation histories but very interesting for studying nucleosynthesis in old stellar populations, such as dSphs \citep{ji_r-process_2016, tsujimoto_enrichment_2017, duggan_neutron_2018}. Comparing the $r$-process element abundances, such as that of Eu, to elements produced on known timescales (such as Mg from CCSNe or Fe from type Ia supernovae) can provide insight into the timescale on which $r$-process elements are produced \citep{hill_vltflames_2019, skuladottir_neutron-capture_2019, skuladottir_evidence_2020}. This timescale is important for uncovering where the $r$-process occurs.

In the $s$-process, neutron capture occurs slowly compared to the $\beta$-decay timescale. The $s$-process primarily occurs in low- to intermediate-mass ($\sim 1$--8~$M_{\sun}$) asymptotic giant branch (AGB) stars  \citep{busso_nucleosynthesis_2001, lugaro_s-process_2003, cristallo_evolution_2011} and potentially in limited amounts in fast-rotating massive stars \citep{pignatari_s-process_2008, frischknecht_non-standard_2012, frischknecht_s-process_2016, choplin_non-standard_2018}, especially pertinent at low metallicities when stars have yet to evolve onto the AGB (around ${\rm [Fe/H]} < -1.5$ for the galaxies in this study). S-process elements are released at the end of AGB stars' lives. Since the evolution of AGB stars is well understood, $s$-process elements have a well-known distribution of delay times, the time between star formation and when those elements are produced. In this case, it is essentially the lifetime of the $\sim 1$--8~$M_{\sun}$ stars that will reach the AGB\@.  How precisely the delay time of an element is known determines how precisely the galaxy's SFH can be determined from that element. By measuring the abundances of $s$-process elements in old stars, which are made from gas that has been enriched by prior populations of stars, including AGBs, $s$-process elements can be used as precise chemical clocks to study the evolution of galaxies \citep[e.g.,][]{skuladottir_neutron-capture_2019, casali_chemicalclock_2020, hayden_galah_2022, viscasillasvazquez_2022}.

To use $s$-process elements to estimate the SFH of a galaxy and to study the production timescale of $r$-process elements, we need a large set of neutron-capture element abundance measurements in old stars. Because a single neutron-capture element can be made in both the $s$- and $r$-processes, it is useful to measure abundances of both primarily $r$-process elements (such as Eu) and primarily $s$-process elements (such as Sr, Y, and Ba). The ratio of two different neutron-capture elements' abundances is a proxy for the $r/s$-process ratio, from which the $s$- or $r$-process-only component of each element may be inferred. 

Large sets of neutron-capture element abundances can be hard to assemble; such elements are typically measured from high-resolution spectra, which are time-intensive to obtain and can only use the brightest stars. Even so, many studies have used neutron-capture abundances in dSphs from high-resolution spectra to study the $r$- and $s$-processes. For example, \citet{skuladottir_neutron-capture_2019} argued that Ba and Eu abundances in Sculptor show $r$-process enrichment at early times in Sculptor's chemical evolution history, implying that neutron star mergers are not the only source of the $r$-process because they would not yet have occurred early in the galaxy's history. \citet{reichert_neutron-capture_2020} corroborated this finding with their reanalysis of Ba and Eu abundances in 13 classical and ultra-faint dwarf spheroidal galaxies. In addition, they found that the onset of the $s$-process in a galaxy increases with metallicity as a function of that galaxy's stellar mass. Other high-resolution neutron-capture abundance studies of dwarf spheroidal galaxies include \citet{shetrone_vltuves_2003, geisler_sculptor-ing_2005, cohen_chemical_2009, cohen_chemical_2010, tsujimoto_enrichment_2017, hill_vltflames_2019, skuladottir_neutron-capture_2020}; and \citet{theler_chemical_2020}, though this is not a comprehensive list. While such studies have significantly improved our understanding of heavy-element nucleosynthesis in dSphs, a larger, homogeneous dataset will facilitate better comparisons with models, probe a larger metallicity range, and greatly increase the number of abundance measurements in the smaller dSphs.

Medium-resolution spectra offer an avenue for assembling a larger sample of neutron-capture abundances in dwarf galaxies because they can target fainter stars. \citet{duggan_neutron_2018} used medium-resolution spectra to measure the abundance of a neutron-capture element, Ba, for the first time. This study demonstrated that medium-resolution spectra can be used to measure neutron-capture abundances on a large scale.

In this paper, we present measurements of Sr, Y, Ba, and Eu in over 500 stars in the Sculptor, Fornax, Draco, Ursa Minor, and Sextans dSphs.  These well-studied galaxies are visible from the northern hemisphere and contain enough stars to make multi-object spectroscopy worthwhile. They span about two orders of magnitude in luminosity and one order of magnitude in mean metallicity.  Their stellar masses range from $5\times 10^5~M_{\sun}$ (Ursa Minor) to $4 \times 10^7~M_{\sun}$ (Fornax) \citep[converting to masses luminosities from][assuming $M_*/L = 2$]{Irwin1995}. 225 of the Ba measurements are a reanalysis of the stars in \citet{duggan_neutron_2018}, while 363 are new measurements of Ba. We also present new Mg measurements for each star. We used medium-resolution spectra from the DEIMOS spectrograph on the Keck~II telescope to assemble the largest homogeneous set of neutron-capture abundance measurements in dwarf galaxies.

In Section \ref{sec:data}, we describe our observations. We outline our method for measuring neutron-capture abundances from the DEIMOS spectra and validate our results in Section \ref{sec:abundances}. In Section \ref{subsec:syserror} we use an identical analysis technique on stars in the globular clusters NGC\,1904, NGC\,2419, NGC\,5024, and NGC\,5904 to estimate the systematic error in our measurements. In Sections \ref{sec:results} and  \ref{sec:discussion}, we place our results in the context of star formation histories and the timescales of nucleosynthetic events and discuss their implications for the $s$- and $r$-processes. We summarize our conclusions in Section \ref{sec:conclusion}.

\section{Observations} \label{sec:data}

\subsection{Data acquisition}

We obtained multi-slit spectra of five dwarf galaxies with Keck/DEIMOS \citep{faber_deimos_2003}.  Observations for all but two of the slitmasks used in this work have been previously published in other analyses \citep{kirby_multi-element_2010,duggan_neutron_2018,delosreyes_manganese_2020}.  Table~\ref{tab:obs} summarizes the observations.

\begin{deluxetable*}{llccc}
\tablecolumns{5}
\tablewidth{0pt}
\tablecaption{DEIMOS Observations\label{tab:obs}}
\tablehead{
 \colhead{System} & \colhead{Slitmask} & \colhead{Grating} & \colhead{UT Date} & \colhead{Exp.\ Time (min.)}}
\startdata
\cutinhead{Globular Clusters}
NGC\,1904 (M79) & ng1904  & 1200B & 2018 Mar 19  & \phn24 \\
NGC\,2419       & n2419b  & 900ZD & 2012 Mar 19 & \phn45 \\
NGC\,5024 (M53) & n5024b  & 1200B & 2019 Mar 10--11 & 107 \\
NGC\,5904 (M5)  & n5904b  & 1200B & 2019 Mar 11 & \phn71 \\
\cutinhead{Dwarf Galaxies}
Sculptor   & scl1  & 900ZD & 2011 Jul 31 & \phn95 \\
           & scl2  & 900ZD & 2011 Aug 6  & \phn68 \\
           & scl5  & 1200B & 2018 Aug 14/2018 Sep 10--11 & 239 \\
           & scl6  & 900ZD & 2011 Aug 4  & \phn81 \\
Fornax     & for6  & 900ZD & 2011 Aug 5--7 & 106 \\
           & for7  & 1200B & 2018 Aug 14/2018 Sep 10--11 & 480 \\
Sextans    & sex2  & 900ZD & 2016 Jan 30--31 & 137 \\
           & sex3  & 900ZD & 2016 Jan 29 & 154 \\
           & sex10\tablenotemark{a} & 1200B & 2021 Apr 9  & 202 \\
Ursa Minor & umi1  & 900ZD & 2011 Jul 29 & \phn90 \\
           & umi2  & 900ZD & 2011 Jul 31 & \phn80 \\
           & umi3  & 900ZD & 2011 Aug 4  & \phn80 \\
           & umia\tablenotemark{a}  & 1200B & 2019 Mar 12/2020 Jul 18/2021 Apr 10 & 460 \\
Draco      & dra1  & 900ZD & 2011 Jul 30 & 100 \\
           & dra2  & 900ZD & 2011 Aug 7  & \phn80 \\
           & dra3  & 900ZD & 2011 Aug 5  & 100 \\
           & dra10\tablenotemark{a} & 1200B & 2021 Apr 9-10 & 203 \\
\enddata
\tablenotetext{a}{These slitmasks were created from the member stars from more than one slitmask previously observed with 1200G \citep{kirby_multi-element_2010}.}
\tablerefs{900ZD spectra were published by \citet{kirby_carbon_2015} and \citet{duggan_neutron_2018}.  The 1200B spectra for scl5, for7, and the 2019 observation of umia were published by \citet{delosreyes_manganese_2020}.  The remaining 1200B spectra are new in this work.}
\end{deluxetable*}

All of the targets in our study were observed by \citet{kirby_multi-element_2010} with the 1200G grating in DEIMOS at a central wavelength of 7800~\AA\@.  This configuration provides an approximate wavelength coverage of 6400--9000~\AA, with variation depending on the location of the slit on the focal plane.  The grating has a groove spacing of 1200 mm$^{-1}$ and a blaze wavelength of 7760~\AA\@.  It provides an approximate resolving power of $R \sim 7000$ when using slits that are 0.7'' wide.

In addition, all of the targets were observed with one of two configurations of DEIMOS that provide bluer spectral coverage.  The first blue configuration, employed by \citet{kirby_carbon_2015} (all dwarf galaxies except Sextans) and \citet{duggan_neutron_2018} (Sextans), used the 900ZD grating, which has a groove spacing of 900 mm$^{-1}$ for a resolving power of approximately $R \sim 2550$ with 0.8'' wide slits.  The blaze wavelength is 5500~\AA, which is also the central wavelength.  The order-blocking filter GG400 protected the spectra from second-order contamination.  The approximate wavelength range was 4000--7200~\AA\@.  The second blue configuration, used by \citet{delosreyes_manganese_2020}, used the newer 1200B grating with a groove spacing of 1200 mm$^{-1}$ and a blaze wavelength of 4500~\AA\@.  The resolving power is approximately $R \sim 3400$.  The central wavelength was 5500~\AA\ for an approximate spectral coverage of 3800--6100~\AA\@.  The central wavelength was chosen to be significantly redder than the blaze wavelength because the optics in DEIMOS become nearly opaque at wavelengths shorter than $\sim 3900$~\AA\@.

Calibration lamps were observed through each slitmask.  A quartz lamp was used for flat fielding.  The blue spectra included two sets of quartz lamp exposures with separate exposure times.  The short exposures were used for the red half of the detector mosaic, and the long exposures were used for the blue half.  Ne, Ar, Kr, Xe, and Hg arc lamps were used for wavelength calibration.  

Most of the slitmasks were originally designed by \citet{kirby_multi-element_2010}, who described the target selection in detail.  In summary, the DEIMOS slitmask design software maximizes ``priority'' while satisfying physical design constraints.  Most importantly, targets must fall on the illuminated area of the slitmask, and their spectra must not overlap with other targets. \citet{kirby_multi-element_2010}\ selected targets from the color--magnitude diagram of each the five dwarf galaxies.  Stars with colors and magnitudes consistent with the red giant branch (RGB) of each galaxy were given high priority.  Within the RGB, brighter stars were given even higher priority.  The lowest priority was assigned to stars that did not lie on the RGB\@.  

Three slitmasks (\texttt{sex10}, \texttt{umia}, and \texttt{dra10}) were designed after \citeauthor{kirby_multi-element_2010}'s (\citeyear{kirby_multi-element_2010}) observations.  Almost all of the targets on those masks were stars identified as members by \citet{kirby_multi-element_2010}.  However, known non-members were not included to make space for new candidate members.  These new candidate members would not have been previously observed with the 1200G grating, which we require for measurements of stellar parameters, such as effective temperature and metallicity.  Therefore, those new stars are not analyzed in this work.

\citet{kirby_multi-element_2010}\ also outlined the procedure for selecting member stars.  Briefly, stars must have colors, magnitudes, radial velocities, and Gaia proper motions \citep{Gaia_EDR3_2021, GaiaDR3_2023} consistent with membership in each galaxy.  In this work, we consider only member stars.  Non-member stars are not analyzed or discussed.

\subsection{Data reduction}

\citet{kirby_multi-element_2010}, \citet{duggan_neutron_2018}, and \citet{delosreyes_manganese_2020} described the procedures for data reduction.  In summary, the spectra were reduced with a modified version of \texttt{spec2d} \citep{cooper_spec2d_2012,newman_deep2_2013}, an IDL code designed for DEIMOS data reduction.  The code provides wavelength-calibrated, flat-fielded, sky-subtracted one-dimensional spectra.  The wavelength calibration included an improved version of emission line selection and rejection \citep{delosreyes_manganese_2020}, and the extraction of 1D spectra from 2D spectra included a correction for differential atmospheric refraction along the slit \citep{kirby_carbon_2015}. The analysis in this paper is carried out on stacked exposures for each star when multiple are available.

The 900ZD and 1200B spectra suffer from ghosting, where stray reflections overlap the stellar spectra.  Ghosting most often appears as distorted, sometimes enlarged images of bright emission lines, such as sky lines.  It also affects the arc lamp calibration images. We improved the identification of arc lines to avoid spurious identification of ghosts.  The improved identification was enabled by taking arc lamp exposures one at a time rather than a single exposure of all lamps.  In that way, the emission lines can be unambiguously associated with their corresponding element.  However, the stellar spectra still sometimes suffer from ghosting.  We mitigate the effect of ghosting by disregarding absorption lines that suffer from very poor model fits (Section~\ref{subsec:MOOG}).

Our neutron-capture abundance measurements start from the known atmospheric parameters (effective temperature $T_{\rm eff}$, surface gravity $\log g$, metallicity [Fe/H], [C/Fe] ratio, [$\alpha$/Fe] ratio) for each star.  These measurements were taken from the catalogs of \citet{kirby_multi-element_2010} for dSphs and \citet{kirby_lithium-rich_2016} for globular clusters, as were radial velocities for each star. They fit the 1200G spectra for each slitmask to a large grid of model spectra \citep[made publicly available by][]{kirby_grids_2011}.  The stellar parameters for each star were taken to be those of the best-fitting, interpolated model spectrum. [C/Fe] ratios were taken as known parameters when available from \citet{kirby_carbon_2015}. We do not propagate the random errors in these atmospheric parameters into our abundance measurements, but we capture these errors in our systematic error estimates (Section \ref{subsec:syserror}.



\section{Abundance Measurements} \label{sec:abundances}

\begin{figure*}[t]
    \centering
    \includegraphics[width=1\textwidth]{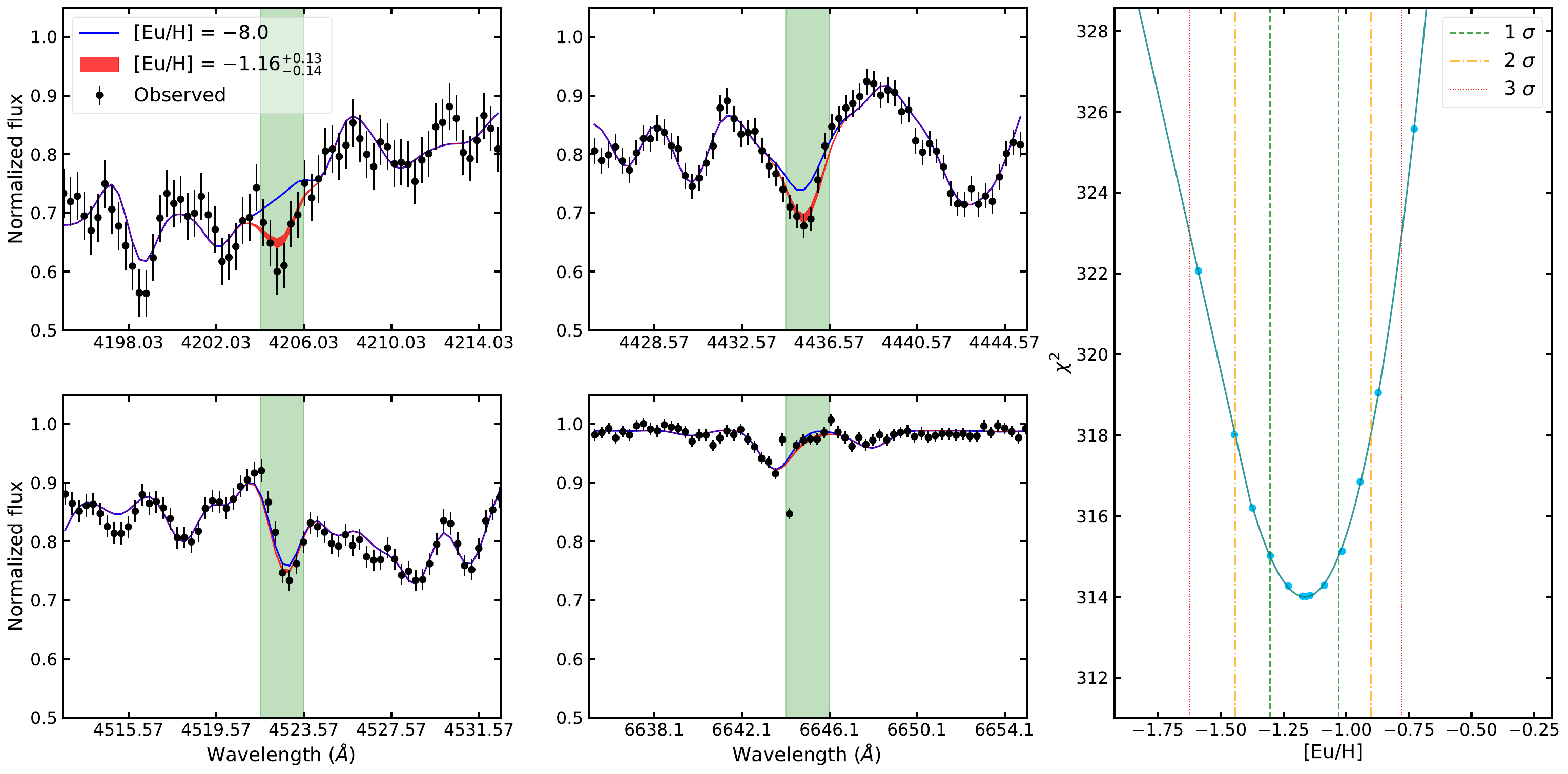}
    \caption{The Eu abundance measurement for the star 1007750 in Sculptor. The best-fit, [Eu/H] = ${-1.16}^{+0.13}_{-0.14}$ (error here does not include $\delta_{\mathrm{sys}}$, see Section \ref{subsec:syserror}), is shown by the red synthetic spectrum. Each of the four left panels is a 20 \AA\ region around a Eu absorption line, with the Eu line highlighted in green. The blue synthetic spectrum has all the same stellar parameters as the red best-fit spectrum, except it has arbitrarily low Eu. The right panel is the $\chi^2$ curve for this abundance measurement. The vertical lines show the 1, 2, and 3 $\sigma$ [Eu/H] values. More discussion of the error determination can be found in Section \ref{subsec:error}.} 
    \label{fig:synth_fit}
\end{figure*}

\begin{figure*}
    \centering
    \includegraphics[width=1\textwidth]{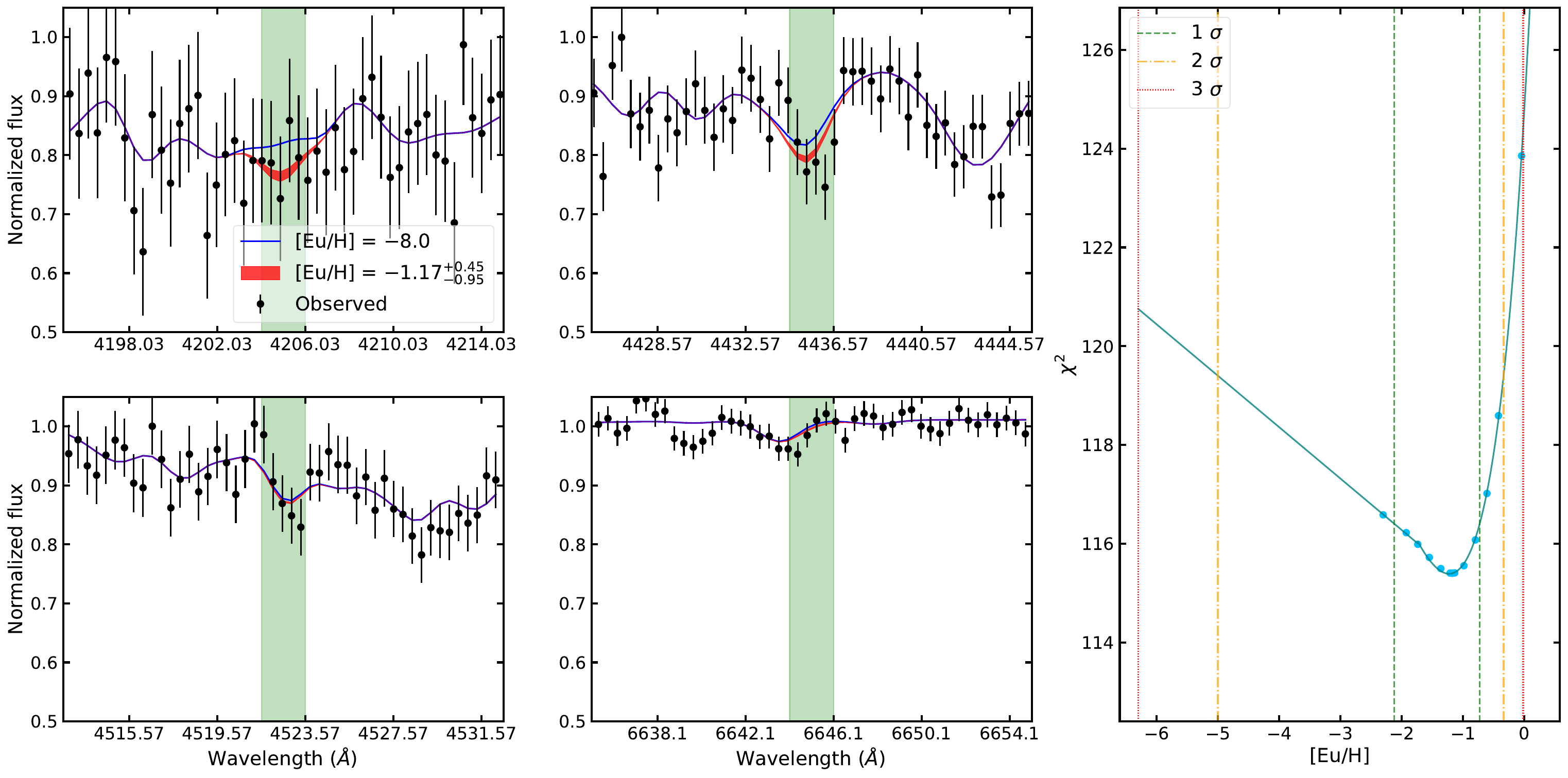}
    \caption{The Eu abundance upper limit determination for the star 1006544 in Sculptor. As in Figure \ref{fig:synth_fit}, the red model spectrum is the best fit with [Eu/H] = ${-1.17}^{+0.45}_{-0.95}$ (error here does not include $\delta_{\mathrm{sys}}$, see Section \ref{subsec:syserror}; the red model spectrum does not span the entire error range as determined from the $\chi^2$ analysis in Section \ref{subsec:error}). Each of the four left panels is a 20 \AA\ region around a Eu absorption line, with the Eu line highlighted in green. The blue synthetic spectrum has the same stellar parameters as the red best-fit spectrum, except it has arbitrarily low Eu. The right panel is the $\chi^2$ curve for this abundance measurement. The vertical lines show the 1, 2, and 3 $\sigma$ [Eu/H] values. Based on the criteria in Section \ref{subsec:error} this measurement is an upper limit. The upper limit reported in this work is the 2 $\sigma$ value at [Eu/H] = -0.33.}
    \label{fig:upperlim_fit}
\end{figure*}

In this section, we describe our method of measuring abundances from the spectra described above. The crux of our method is that we fit a synthetic spectrum generated by \texttt{MOOG} \citep{sneden_MOOG_2012} to each of the lines from the element of interest simultaneously, allowing us to take advantage of nearly all the information about the element in the spectrum. This paper uses an updated and generalized form of the SMAUG code described by \citet{delosreyes_manganese_2020} \citep{smaug20}.
\footnote{SMAUG code used in \citet{delosreyes_manganese_2020}: \url{https://github.com/mdlreyes/SMAUG}}. We summarize the key features of the code and the changes we have made. The abundance analysis proceeds as follows:
\begin{enumerate}
    \item Continuum-normalize the spectra.
    \item Adopt $T_{\rm eff}$, $\log g$, [Fe/H], [$\alpha$/Fe], and [C/Fe] from prior measurements (see Section \ref{sec:data}).
    \item Use \texttt{MOOG} to model 20 \AA\ regions around the absorption lines of one neutron-capture element.
    \item Adjust the abundance of the neutron-capture element to find the best-fitting model spectrum.
    \item Exclude lines of the element of interest that have a notably worse fit to the observed spectrum due to ghosting or poor wavelength calibration. Perform step 4 again.
\end{enumerate}

\noindent
The following subsections detail these steps.

\subsection{Spectrum Preparation} \label{subsec:specprep}
Our spectra are not flux-calibrated, and the spectral models are continuum-normalized. Therefore, we must normalize the observed spectrum before measuring abundances. As in the work by \citet{delosreyes_manganese_2020}, we first interpolate grids of pre-generated continuum-normalized spectra, generated with \texttt{MOOG}  \citep{kirby_grids_2011,kirby_carbon_2015, escala_elemental_2019}, to the proper stellar parameters \citep{kirby_multi-element_2010,kirby_carbon_2015,kirby_lithium-rich_2016}. The grid spectra have varying Fe, C, and $\alpha$ abundances, and the solar abundance ratio of other elements. A Gaussian kernel is used to smooth the synthetic spectrum from the grid to match the resolution of the observed spectrum. 

The synthetic spectrum is then used to determine the continuum as follows. We divide the observed spectrum by the synthetic spectrum, with regions that have significant variations from the continuum (Balmer lines, Na~D doublet, around the CCD chip gap) masked out. We also masked out 20~\AA\  regions centered on the lines of the element of interest. A third-order spline is fit to the quotient and the observed spectrum is divided by this spline, which represents the continuum shape of the spectrum. This leaves a continuum-normalized spectrum that can be used to measure abundances.

We also applied small corrections to the wavelength solutions of the spectra. We did this by finding strong lines in the spectrum (Balmer series, Mg, and Na~D lines) and calculating the difference between the observed wavelength of the line and its true rest wavelength. We fit a line to those shifts vs.\ wavelength for each side of the chip gap and multiplied the spectrum by that linear shift. This method allows for different corrections at different wavelengths by slightly stretching/compressing the spectrum where needed. Typically this was a small difference, but it improved the fits at the red end of the spectrum, which was especially important for Mg, Ba, and Eu.

\begin{deluxetable}{cccc}
\tablecolumns{4}
\tablewidth{0pt}
\tablecaption{Line List\label{tab:elemlines}}
\tablehead{
 \colhead{Species} & \colhead{Wavelength (\AA)} & \colhead{Excitation Potential (eV)} & \colhead{log gf}}
\startdata
Mg I & 4571.096 & 0.00  & -5.620 \\
Mg I & 4702.991 & 4.346 & -0.440 \\
Mg I & 5167.322 & 2.709 & -0.870 \\
Mg I & 5172.684 & 2.712 & -0.390 \\
Mg I & 5183.604 & 2.717 & -0.170 \\
Mg I & 5528.405 & 4.346 & -0.500 \\
Mg I & 5711.088 & 4.343 & -1.830 \\
\hline
Sr II & 4215.524 & 0.000 & -0.170 \\
Sr I & 4607.331 & 0.000 & 0.280 \\ 
\hline
Y II & 4398.010 & 0.129 & -1.000 \\
Y II & 4854.867 & 0.992 & -0.380 \\
Y II & 4883.680 & 1.080 & 0.070 \\
Y II & 4900.110 & 1.032 & -0.090 \\
Y II & 5087.420 & 1.083 & -0.170 \\
Y II & 5200.413 & 0.992 & -0.570 \\
Y II & 5205.731 & 1.032 & -0.340 \\
\hline
Ba II & 4554.034 & 0.000 & 0.170 \\
Ba II & 4934.100 & 0.000 & -0.157 \\
Ba II & 5853.690 & 0.604 & -1.010 \\
Ba II & 6141.730 & 0.704 & -0.077 \\
Ba II & 6496.910 & 0.604 & -0.380 \\
\hline
Eu II & 4205.026 & 0.000 & 0.210 \\
Eu II & 4435.568 & 0.207 & -0.110 \\
Eu II & 4522.573 & 0.207 & -0.670 \\
Eu II & 6645.104 & 1.379 & 0.120 \\
\enddata
\end{deluxetable}

\subsection{Inputs for Abundance Measurements} \label{subsec:inputs}
The abundance measurements depend on two main inputs: line lists of atomic data (wavelength, excitation potential, and oscillator strength) and physical parameters of the star (temperature, surface gravity, and metallicity). Spectral synthesis also requires stellar atmosphere models, for which we used \texttt{ATLAS9} model atmospheres \citep{kurucz_new_1993,kirby_grids_2011}.

The atomic lines used to measure each element presented in this paper are well-studied absorption lines that are often used in high-resolution spectroscopic abundance studies (in particular, we referred to \citealt{ji_s5_2020}). The list of lines we used for each element can be found in Table \ref{tab:elemlines}. We used \texttt{linemake} \citep{placco_linemake_2021} to construct a line list for a 20 \AA\ region around each line.

Line splitting due to hyperfine structure and isotopic splitting can become important for heavy elements. We included isotope ratios from \citet{sneden_neutron-capture_2008} in the measurements for Ba and Eu. These isotope ratios may not be perfect for the stars in this study, as they are based on solar abundances. Nevertheless, they provide a good first approximation. The linelists from \texttt{linemake} specify isotopes in their molecular lines as well, so we included isotope ratios for $\mathrm{{}^{12}C}$ and $\mathrm{{}^{13}C}$ following the $\mathrm{{}^{12}C/{}^{13}C}$ relation with $\log g$ in \citet{kirby_carbon_2015} (which was adapted from \citealt{keller_12c13c_2001}) for the CH and CN lines:
\begin{equation}
    \begin{aligned}
        \mathrm{{}^{12}C/{}^{13}C} = 50 \text{  if $\log g$ $>$ 2.7}\\
        \mathrm{{}^{12}C/{}^{13}C} = 63\: \text{$\log g$} - 120 \text{  if 2.0 $<$ $\log g$ $\leq$ 2.7}\\
        \mathrm{{}^{12}C/{}^{13}C} = 6 \text{  if $\log g$ $\leq$ 2.0} \label{eq:carbonratio}
    \end{aligned}
\end{equation}

\noindent
The scatter in this relation is large, but the isotopic ratio has a minor influence on the spectra at DEIMOS resolution. For MgH, we assume solar Mg isotope ratios from \citet{asplund_chemical_2009}.

Some of the elements we measured have lines in the carbon molecular bands (CH and CN) from $\sim$4100--4400~\AA\@. The evolved stars in this study have depleted C abundances, so the solar carbon abundance ratio would be a poor choice in our models \citep{placco_carbon-enhanced_2014}. We adopted the [C/Fe] measurements of \citet{kirby_carbon_2015}, who fit the same spectra that we use here to model spectral grids of the carbon G band. For stars that did not have carbon measurements available, we used the empirical formula for [C/Fe] as a function of stellar luminosity (calculated from $\mathrm{T_{eff}}$, $\log g$, and the assumption of $M = 0.75 M_\odot$, a typical mass of an RGB star in this study) from \citet{kirby_carbon_2015}:
\begin{equation}
    \begin{aligned}
        \mathrm{[C/Fe]} = 0.12 \text{  if log(L/${L}_\odot$) $\leq$ 1.6}\\
        \mathrm{[C/Fe]} = 1.42 - 0.82\text{ log(L/${L}_\odot$)  if log(L/${L}_\odot$) $>$ 1.6} \label{eq:carbon}
    \end{aligned}
\end{equation}

\noindent The scatter in this relation is about 0.2~dex.  Neutron-capture absorption lines in the G-band could have a systematic error from this relation by up to 0.2~dex.  Fortunately, most neutron-capture elements have at least one line outside of the G-band.

For each star, $T_{\rm eff}$, $\log g$, microturbulent velocity ($\xi$), [Fe/H], [$\alpha$/Fe], and [C/H] are specified when making the model spectra. The value for $\xi$ is calculated using $\log g$ following the empirical relation in \citet{kirby_multi_element_2009}:
\begin{equation}
    \begin{aligned}
        \xi(\mathrm{km\, s^{-1}}) = 2.13 - 0.23\text{ $\log g$} \label{eq:turbulence}
    \end{aligned}
\end{equation}

\noindent The typical uncertainty on this relation is 0.06~km~s$^{-1}$, which gives an error of less than 0.1~dex in derived abundances.  Therefore, the error in this relation is a minor contributor to the error budget.

We assume ${\rm [X/Fe]} = 0$ for elements other than carbon, the $\alpha$ elements (in this case, Mg, Si, S, Ar, Ca, and Ti based on [$\alpha$/Fe] from \citet{kirby_multi-element_2010}), and the element we are measuring. In addition to the neutron-capture elements we measured, we also used this technique to measure [Mg/Fe] for each star. [Mg/Fe] measurements are valuable because they allow for comparisons with a prompt delay time, as Mg is mostly made in CCSNe. Although \citet{kirby_multi-element_2010} previously measured Mg abundances from spectra using the DEIMOS 1200G grating, those spectra did not include the Mg~b triplet.  As a result, their Mg abundance measurements are based on weak lines and have large uncertainties.  Our bluer spectra do include Mg~b, which is very effective for measuring Mg abundances.

\begin{figure*}
    \centering
    \includegraphics[width=1\linewidth]{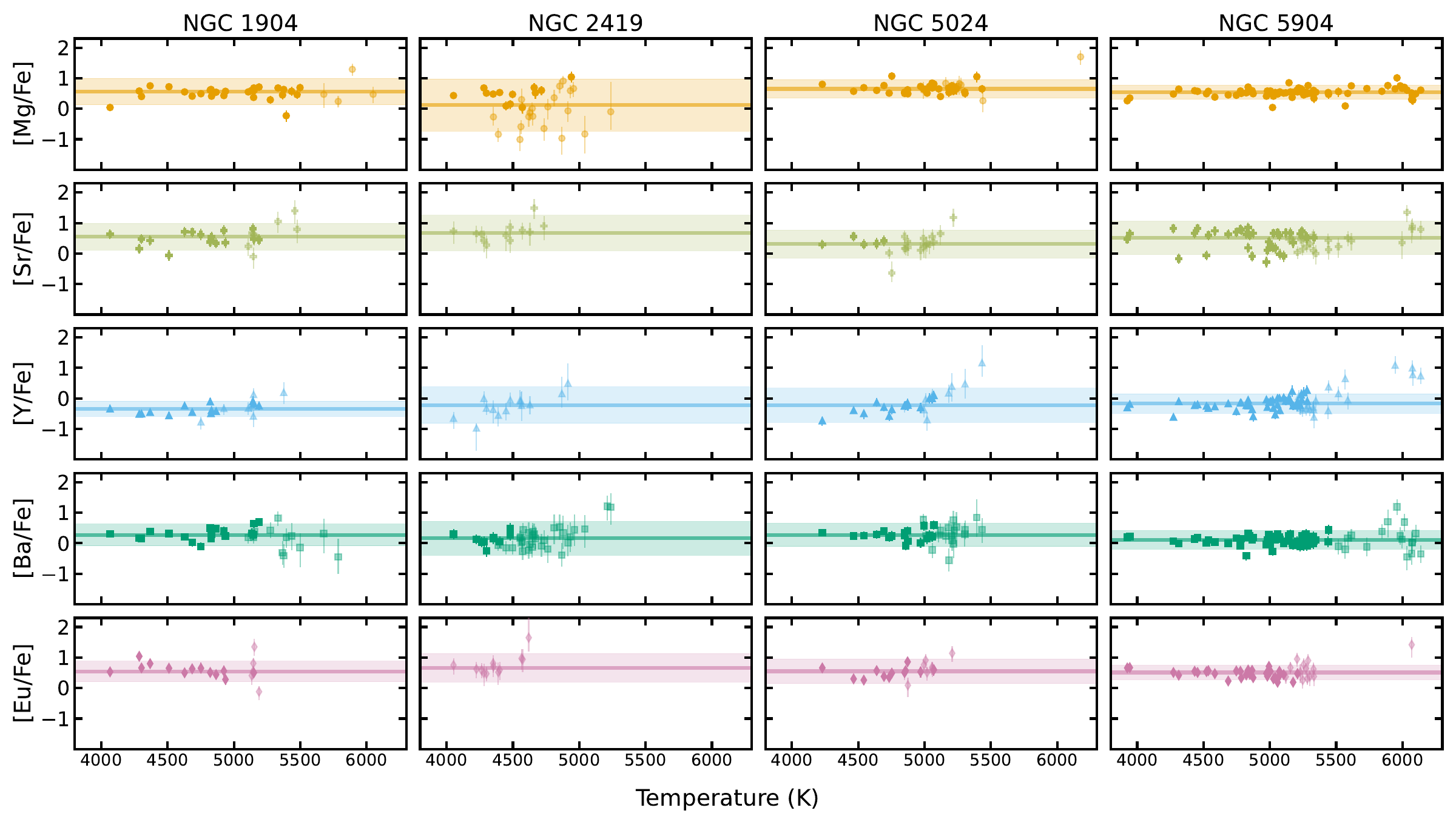}
    \caption{Abundance measurements for NGC\,1904, NGC\,2419, NGC\,5024, and NGC\,5904 as a function of stellar $T_{\rm eff}$. Stars with error $>$ 0.3 dex, including systematic error derived in Section \ref{subsec:syserror}, are less opaque. The median in each panel is plotted as a solid line, with a 2 $\sigma$ region around the median lightly shaded.
    \label{fig:temperature}}
\end{figure*}

\subsection{Spectrum modeling} \label{subsec:MOOG}

To reduce computation time, we synthesize only 20~\AA\ regions around each line of interest (the lines used are listed in Table \ref{tab:elemlines}). The best-fit abundance of the element of interest is found by weighted least-squares fitting. \citet{delosreyes_manganese_2020} provide more information on this procedure.

To ensure that the synthetic spectrum was a good match to the normalized flux of the observed spectrum, we corrected for any overall continuum offset in the spectra during the abundance measurement. For each spectral region synthesized by \texttt{MOOG}, we divided the observed flux by the synthetic flux, fit a line to that quotient, and multiplied the synthetic spectrum by that line. This process had little effect for most lines, but was useful for lines in or near the G band due to blanketing by CH and CN lines. The \ion{Sr}{2} line at 4216 \AA\ is one of the lines that most benefited from local continuum normalization. For example, if our estimate of [C/H] was too low, the synthetic spectrum would lie above the observed spectrum and the [Sr/H] measurement would be artificially large to compensate for the local continuum difference. This extra step ensured that any inconsistencies in [C/H], the modeling of molecular lines, and the continuum normalization had a minimal impact on our abundances.

In order to produce reliable measurements from as many stars as possible, we run the least-squares fit to the element of interest twice for some stars. The first time proceeds as described above. The second time, any 20 \AA\ region around a line whose model is a very poor fit to the observed spectrum is removed before running again. This way, areas of the synthetic spectrum that do not fit the observations (due to ghosting, wavelength calibration issues, or poor models) do not skew the abundance measurement. Some stars, especially in the globular clusters, have such low error on the observed spectrum that differences between the model and the observed spectrum are dominated by systematic error. This makes reduced $\chi_{\rm red}^2$ an unideal quantity for assessing the goodness of fit on its own. To compensate for this, we only remove a line from the analysis if the fit to the 20 \AA\ region around that line has a significantly worse $\chi_{\rm red}^2$ than the other lines for that same element, based on the median $\chi_{\rm red, median}^2$. The lines that are left out due to an unideal model fit have $\chi_{\rm red}^2$ $>$ $\chi_{\rm red,median}^2 + 2\sigma$ and $\chi_{\rm red}^2$ $>$ 5. For the stars that undergo this iterative method, the abundance measurements rely on the information from fewer lines than for other stars (in addition to any lines missing due to differences in wavelength range between the spectra).

\subsection{Validation and Error Determination} \label{subsec:error}
The least-squares fit to the observed spectrum described above produces an error estimate for the model fit. However, describing the error in our abundance measurements with one number does not reflect the fact that the error is likely asymmetric. At lower [X/H] values, the model spectrum is less responsive to changes in abundance. For many stars, this results in a larger error bar below the [X/H] measurement than above it.

To calculate asymmetric error and to determine which measurements are upper limits, we calculated the total $\chi^2$ for the model fit to all the lines for a range of [X/H] values around the best fit. We fit a curve to the $\chi^2$ values for each abundance measurement.  Figure \ref{fig:synth_fit} includes an example of this process. The $\chi^2$ values were fit with a parabola when symmetric, or, more often, a function of the form $e^{ax+d}-b(x+d)+c$, where one side (in this case to lower values of [X/H]) can have a lower absolute slope than the other. This reproduces the fact that changes in [X/H] at low values of [X/H] produce little to no change in the model spectrum fit. We also included a linear fit of the three $\chi^2$ points at the lowest [X/H] values modeled because the model spectra are less sensitive to changes in abundances at low values and those points were not well-fit by the other functions. The $1\sigma$ asymmetric error is calculated by finding the [X/H] value that corresponds to $\chi_{\rm minimum}^2$ + 1 on either side of the best fit. The $2\sigma$ error is found with $\chi_{\rm minimum}^2$ + 4 and so on.

Using these $\chi^2$ curves, we applied some criteria to determine whether a measurement is an upper limit. If the 1$\sigma$ [X/H] value is more than 1 dex from the best fit [X/H] value, if the 2$\sigma$ [X/H] is more than 1 dex from the 1$\sigma$ [X/H], or if the 3$\sigma$ [X/H] value is more than 4 dex away from the lowest [X/H] value modeled for the $\chi^2$ calculations (best-fit $\mathrm{[X/H]}-3\sigma_{\rm least-squares}$) then the measurement is considered an upper limit. These criteria are essentially checking for highly asymmetric $\chi^2$ curves where models with lower [X/H] do not change the goodness of fit to the observations. An example of an upper limit based on these criteria is shown in Figure \ref{fig:upperlim_fit}.

The errors reported for each abundance measurement in Table \ref{tab:abundances} are the asymmetric errors calculated as described above with systematic errors (see Section \ref{subsec:syserror}) added in quadrature. All of the subsequent plots of abundances in this paper include both the asymmetric error and the systematic error.

\begin{deluxetable}{cc}
\tablewidth{0pt}
\tablecaption{Systematic Errors from Globular Clusters\label{tab:syserr}
}
\tablehead{
 \colhead{Abundance} & \colhead{${\delta}_{\rm sys}$}}
\startdata
${\rm [Mg/Fe]}$ & 0.10 \\
${\rm [Sr/Fe]}$ & 0.16 \\
${\rm [Y/Fe]}$ & 0.11 \\
${\rm [Ba/Fe]}$ & 0.12 \\
${\rm [Eu/Fe]}$ & 0.11 \\
\enddata
\end{deluxetable}

\subsection{Systematic error with globular clusters} \label{subsec:syserror}

To estimate systematic error in the abundances of each element, we followed the same procedure to measure neutron-capture elements in four globular clusters: NGC\,1904 (M79), NGC\,2419, NGC\,5024 (M53), and NGC\,5904 (M5).

Globular clusters are typically assumed to be simple stellar populations, having formed from one molecular cloud. Under this assumption, stars in globular clusters should have uniform metallicities and heavy element abundances. We used this property of globular clusters to estimate our systematic error.

We estimate the standard deviation of the abundances of each neutron-capture element in each of the four globular clusters following Equation \ref{eq:syserr}. This equation essentially calculates the weighted standard deviation and subtracts the random error (for which we used the average of the upper and lower asymmetric errors for each measurement; $\delta_{\rm random}$ in Equation \ref{eq:syserr}), leaving us with the systematic error. We include a term to account for sampling bias using the effective number of degrees of freedom following \citet[Ch. 8.2 and 11.7]{kish_survey}. This extra term is important for recovering systematic error when the sample size is small. For every neutron-capture element we average the systematic errors derived from each globular cluster, resulting in the values quoted in Table \ref{tab:syserr}.

\begin{equation}
    \begin{aligned}
        \mathrm{w = \frac{1}{\delta_{random}^2}} \\[2 ex]
        \mathrm{\mu = \frac{\Sigma w[X/Fe]}{\Sigma w}} \\[2 ex]
        \mathrm{\delta_{sys}^2 = \frac{\Sigma w([X/Fe]-\mu)^2}{\Sigma w - \frac{\Sigma w^2}{\Sigma w}} - \frac{\Sigma w\delta_{random}^2}{\Sigma w}} \label{eq:syserr}
    \end{aligned}
\end{equation}


This method assumes that all the stars in the globular cluster have the same abundances of neutron-capture elements. In reality, some globular clusters have a spread in neutron-capture element abundances (e.g., M15, \citealt{sneden_star--star_1997}; M4, \citealt{marino_ngc6121_2008}; M22, \citealt{marino_ngc6656_2009}; M92, \citealt{kirby_r-process_2023}). Thus, our systematic error is likely an overestimation.

The error quoted for [Mg/Fe] in Table \ref{tab:syserr} is not calculated in the same way. Globular clusters are known to have dispersions in their light element abundances, which includes Mg \citep{gratton_abundance_2004}. Assuming each cluster has a uniform [Mg/Fe] would result in an anomalously high ${\delta}_{\rm sys}$, even more so than for neutron-capture elements. Using the same method as above, we calculate the systematic error in [Mg/Fe] to be 0.19 dex. NGC\,5024 and NGC\,5904 have Mg $\delta_{\rm sys}$ $\approx$ 0.11, while NGC\,1904 and NGC\,2419 bring the average up. This, combined with the fact that the [Mg/Fe] measurements are based on the very strong Mg triplet at $\sim$5170 \AA\@, suggests that choosing a slightly lower error than for the neutron-capture elements is appropriate. In the end, we added a ${\delta}_{\rm sys}$ of 0.10 to the [Mg/Fe] measurements.  

\begin{figure}
    \centering
    \includegraphics[width=1\linewidth]{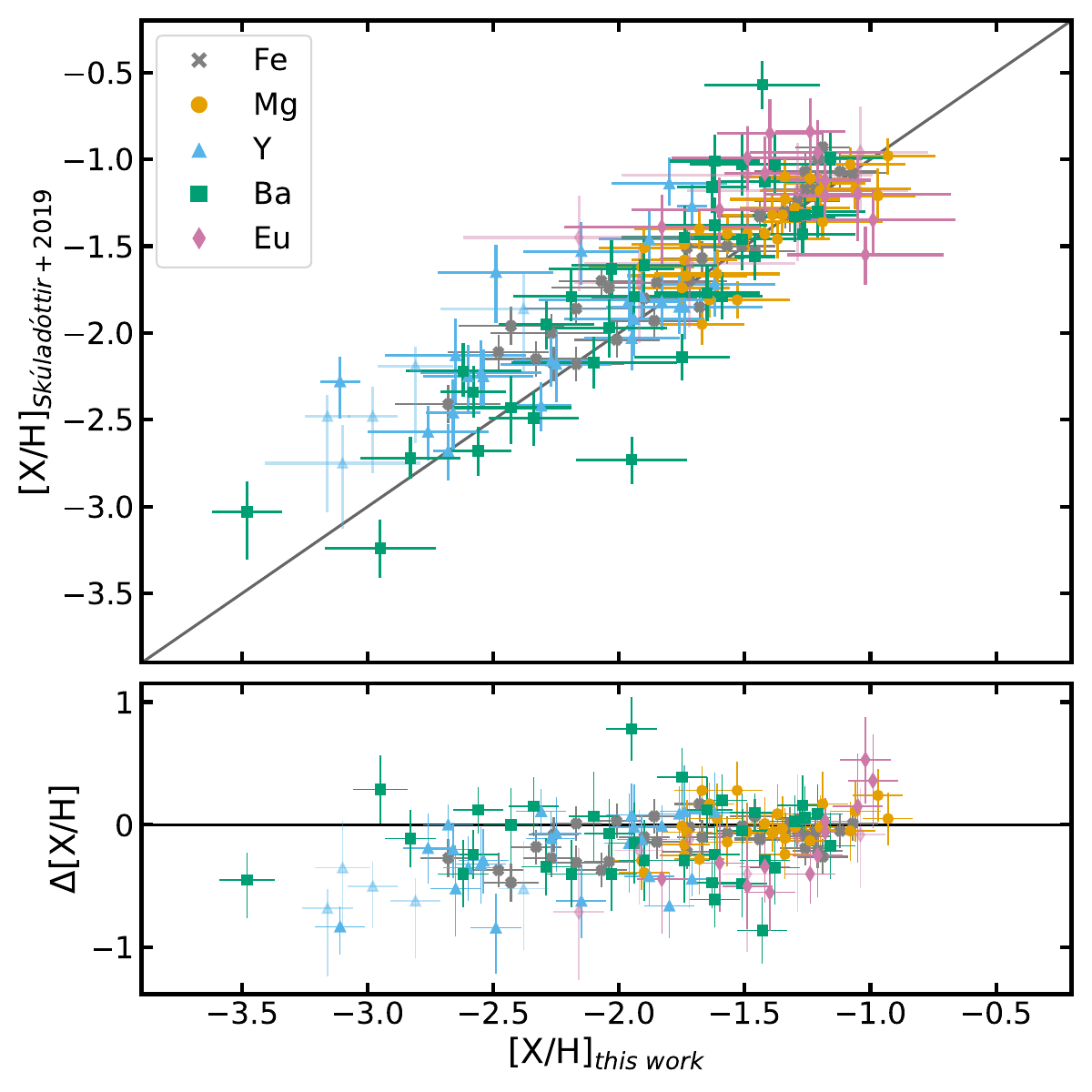}
    \caption{Comparison of Mg, Y, Ba, and Eu abundances from this work and Fe from \citet{kirby_multi-element_2010} with high-resolution measurements from \citet{skuladottir_neutron-capture_2019} for Sculptor. In the lower panel, $\Delta$[X/H] = [X/H]$_{this\ work}$ - [X/H]$_{Sk\acute{u}lad\acute{o}ttir+\ 2019}$. The [X/H] values from \citet{skuladottir_neutron-capture_2019} are scaled from the solar abundances of \citet{grevesse_standard_1998} to those of \citet{asplund_chemical_2009}, which we use in this study. Systematic error derived in Section \ref{subsec:syserror} is included in the error bars on the abundances from this work. Measurements with upper or lower asymmetric errors (including $\delta_{\rm sys}$) $>$ 0.3 are less opaque. The $x=y $ line in the background of the upper panel and the $\Delta$[X/H]=0 in the lower panel are plotted for reference.}
    \label{fig:Sclhigh-res}
\end{figure}

\begin{figure}
    \centering
    \includegraphics[width=1\linewidth]{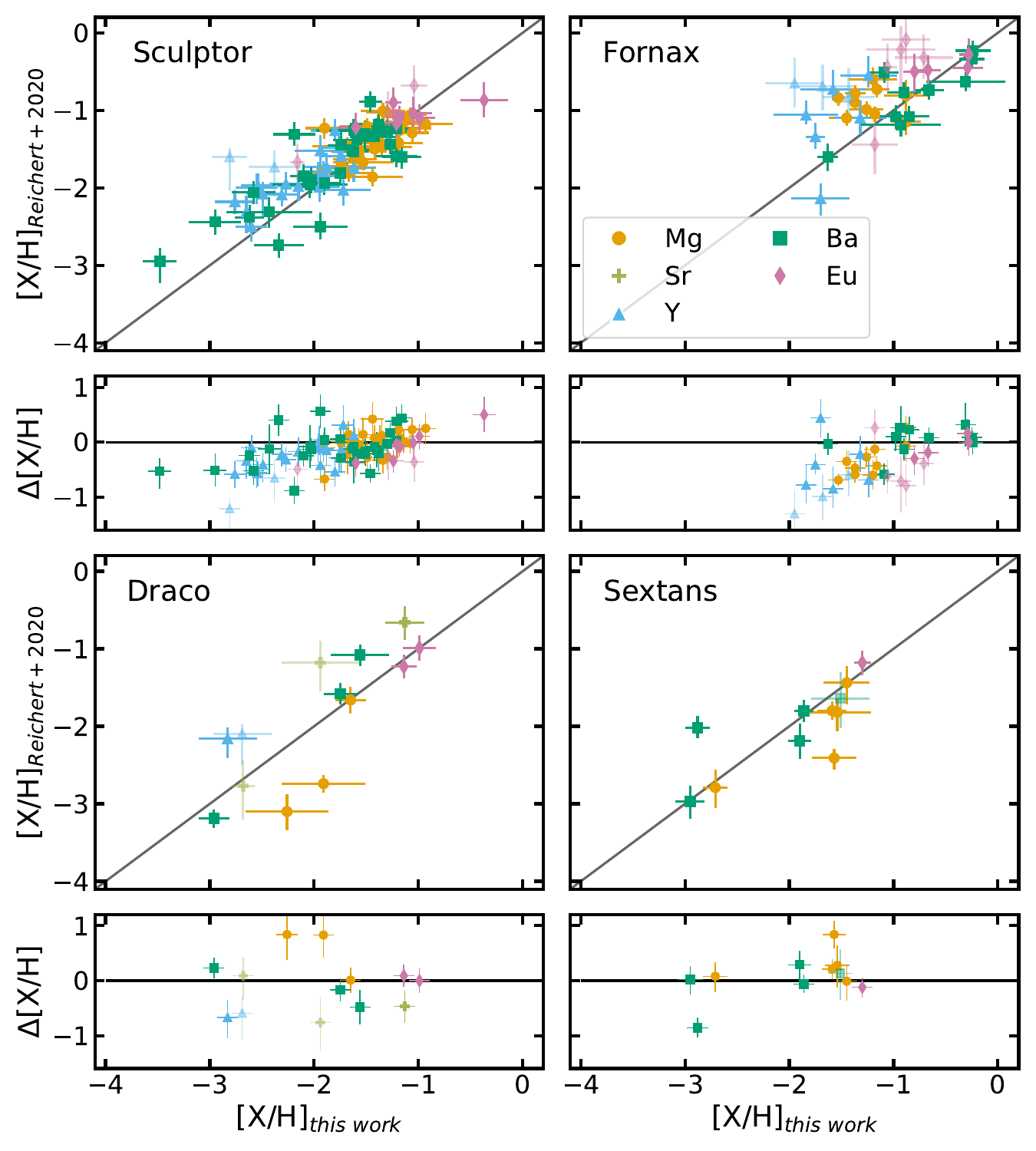}
    \caption{Comparison of Mg, Sr, Y, Ba, and Eu abundances from this work with high-resolution measurements from \citet{reichert_neutron-capture_2020} for Sculptor, Fornax, Draco, and Sextans. In the lower panel, $\Delta$[X/H] = [X/H]$_{this\ work}$ - [X/H]$_{Reichert+\ 2020}$. Systematic error derived in Section \ref{subsec:syserror} is included in the error bars on the abundances from this work. Measurements with upper or lower asymmetric errors (including $\delta_{\rm sys}$) $>$ 0.3 are less opaque. The $x=y$ lines in the background of the larger panels and the $\Delta$[X/H]=0 lines in the offset panels are plotted for reference.}
    \label{fig:Reichert}
\end{figure}

\subsubsection{Local Thermodynamic Equilibrium} \label{subsec:LTE}

We have assumed local thermodynamic equilibrium (LTE) in our abundance measurements, as \texttt{MOOG} and \texttt{ATLAS} are both mostly LTE codes. Some spectral features are sensitive to departures from LTE; corrections for non-local thermodynamic equilibrium (NLTE) effects must be applied in some cases. Here we explore whether that is necessary for our abundance measurements.

\citet{sitnova_cetus_2024} estimated NLTE corrections for Mg, Sr, Y, Ba, and Eu ranging from $\pm$0-0.2 dex for stars with similar temperatures and metallicities to the stars in our sample. \citet{bergemann_nlte_2012} found NLTE corrections to the Sr I line at 4607~\AA\ of about $+$0.4 dex. However, the corrections to the Sr II 4216~\AA\ line were not calculated and their grid does not include $\log g$ $<$ 2.2. Ideally, the NLTE effects of the Sr I line at 4607~\AA\ have been mitigated by our method of fitting both Sr lines simultaneously. The NLTE corrections for Mg in \citet{bergemann_nlte_2017}, Y in \citet{storm_nlte_2023}, and Ba in \citet{korotin_nlte_2015} also fall within this range. In many cases, this is smaller than the random error on a neutron-capture abundance measurement.

Even so, we plotted our abundances as a function of temperature for globular clusters (Figure \ref{fig:temperature}) as a rudimentary test of the effects of the LTE assumption. Because globular clusters have a very narrow metallicity range, they should show fewer true trends with temperature than dwarf galaxies. Figure~\ref{fig:temperature} shows that most of the elements have little to no correlation with temperature. In NGC\,5024 and NGC\,5904, Y abundances begin to increase at higher temperatures but show a simultaneous increase in the error on the abundance measurement. Thus, NLTE effects do not appear to cause significant trends in abundances for the stars in our sample.

\subsection{Comparisons with high-resolution spectroscopy} \label{subsec:comparisons}

To validate our results, we compared our abundance measurements with those from high-resolution spectroscopy where available. Figure \ref{fig:Sclhigh-res} compares our measurements for the subset of our stars in Sculptor that are also in the high-resolution spectroscopic studies of \citet{skuladottir_neutron-capture_2019} and \citet{hill_vltflames_2019}. The high-resolution studies include measurements of Mg, Y, Ba, and Eu. Most of our abundance measurements match well with those from high-resolution spectroscopy within 1$\sigma$.

Our measurements might differ from those of \citet{skuladottir_neutron-capture_2019} for several reasons. One important factor is the difference in lines used for each element. Due to the large wavelength range of our spectra and our choice to synthesize spectra (and therefore use blended lines that are inappropriate for measuring equivalent widths), we used more lines for Mg, Ba, and Eu than were used for the high-resolution measurements. This includes four additional Mg lines (including two more lines in the Mg triplet), two more Ba lines, and three more Eu lines than \citet{skuladottir_neutron-capture_2019}, who only used the Eu line at 6645~\AA\@. Their measurements of Y include two lines not used in our measurements: the singly-ionized Y lines at 5123 and 5509~\AA\@. We decided not to include these lines because they are so weak in the DEIMOS spectra that they were not contributing any significant additional constraint on the Y abundances. 

In Figure \ref{fig:Sclhigh-res}, the [X/H] values from our study are about 0.15 dex less than those of \citet{skuladottir_neutron-capture_2019} on average (including [Fe/H]). We are uncertain what causes this abundance offset, but it does not severely affect our conclusions. First, the offset in [Fe/H] is about the same as for the other elements, so [X/Fe] values are consistent between the datasets and still allow for direct comparison. Second, the power of the measurements presented here from medium-resolution spectra are not in the absolute abundances, for which high-resolution studies may be better-suited, but for relative comparisons between stars in a population. The offsets for each element in Figure~\ref{fig:Sclhigh-res} are roughly constant, so the overall trends observed in [X/Fe] (discussed in Section \ref{sec:discussion}) are not affected.


The abundances presented by \citet{reichert_neutron-capture_2020} offer a point of comparison for some of the other dSphs as well. The results of that comparison can be found in Figure \ref{fig:Reichert}. In Sculptor, the Mg, Eu, and Ba abundances do not exhibit the same offset as in the comparison with \citet{skuladottir_neutron-capture_2019}. This is another reason an overall abundance offset between our measurements and those in particular high-resolution studies is not cause for concern. On the other hand, our [Y/H] measurements appear to be systematically lower than those of \citet{reichert_neutron-capture_2020} in both Sculptor and Fornax. Again, absolute abundances are not our goal. Overall, our abundance measurements from medium-resolution spectra are in good agreement with high-resolution measurements.

\movetabledown=5.8in
\begin{rotatetable}
\begin{deluxetable}{cccccccccccccc}
\tablecaption{Multi-Element Abundance Catalog of dSph Stars \label{tab:abundances}}
\tablehead{\colhead{Name} & \colhead{RA} & \colhead{Dec} & \colhead{slitmask} & \colhead{$T_{\rm eff}$ (K)} & \colhead{$\log g$} & \colhead{[Fe/H]} & \colhead{[$\alpha$/Fe]} & \colhead{[C/Fe]} & \colhead{[Mg/Fe]} & \colhead{[Sr/Fe]} & \colhead{[Y/Fe]} & \colhead{[Ba/Fe]} & \colhead{[Eu/Fe]}}
\startdata
1002447 & $0^{\mathrm{h}}59^{\mathrm{m}}21.66^{\mathrm{s}}$ & $-33^\circ41{}^\prime01.61{}^{\prime\prime}$ & bscl1 & 4709 & 1.27 & -2.02$\pm$0.1 & 0.26 & -0.86 & $+0.67^{+0.11}_{-0.12}$ & $+0.98^{+0.31}_{-0.39}$ & $-0.58^{+0.25}_{-0.30}$ & $-0.52^{+0.26}_{-0.26}$ & $+0.53^{+0.31}_{-0.38}$ \\[2mm]
1002888 & $0^{\mathrm{h}}59^{\mathrm{m}}23.85^{\mathrm{s}}$ & $-33^\circ42{}^\prime59.12{}^{\prime\prime}$ & bscl1 & 4831 & 1.87 & -1.97$\pm$0.12 & 0.11 & -1.14 & $+0.18^{+0.16}_{-0.18}$ &  &  & $-0.66^{+0.42}_{-0.66}$ &  \\[2mm]
1003386 & $0^{\mathrm{h}}59^{\mathrm{m}}26.87^{\mathrm{s}}$ & $-33^\circ40{}^\prime28.90{}^{\prime\prime}$ & bscl1 & 4695 & 1.49 & -1.29$\pm$0.1 & 0.06 & -0.68 & $+0.08^{+0.11}_{-0.12}$ & $<+0.54$ & $-0.49^{+0.32}_{-0.31}$ & $-0.56^{+0.27}_{-0.26}$ & $+0.49^{+0.29}_{-0.34}$ \\[2mm]
1003443 & $0^{\mathrm{h}}59^{\mathrm{m}}27.34^{\mathrm{s}}$ & $-33^\circ38{}^\prime47.34{}^{\prime\prime}$ & bscl1 & 4435 & 1.26 & -1.82$\pm$0.1 & -0.11 & -1.03 &  & $<0.00$ &  & $-0.50^{+0.22}_{-0.23}$ & $+0.40^{+0.25}_{-0.25}$ \\[2mm]
1003505 & $0^{\mathrm{h}}59^{\mathrm{m}}27.13^{\mathrm{s}}$ & $-33^\circ43{}^\prime42.13{}^{\prime\prime}$ & bscl1 & 4391 & 0.88 & -1.81$\pm$0.1 & 0.05 & -1.1 & $+0.42^{+0.10}_{-0.10}$ &  & $-0.65^{+0.22}_{-0.27}$ & $-0.19^{+0.19}_{-0.20}$ & $+0.21^{+0.24}_{-0.33}$ \\[2mm]
1003537 & $0^{\mathrm{h}}59^{\mathrm{m}}27.69^{\mathrm{s}}$ & $-33^\circ40{}^\prime35.41{}^{\prime\prime}$ & bscl1 & 4279 & 0.58 & -2.26$\pm$0.1 & 0.21 & -0.96 & $+0.62^{+0.10}_{-0.10}$ & $+0.41^{+0.28}_{-0.30}$ & $-0.42^{+0.21}_{-0.22}$ & $-0.30^{+0.17}_{-0.18}$ & $+0.03^{+0.26}_{-0.38}$ \\[2mm]
1003694 & $0^{\mathrm{h}}59^{\mathrm{m}}28.30^{\mathrm{s}}$ & $-33^\circ42{}^\prime07.21{}^{\prime\prime}$ & bscl1 & 3920 & 0.32 & -1.68$\pm$0.1 & -0.04 & -1.42 & $+0.29^{+0.13}_{-0.12}$ & $+0.09^{+0.30}_{-0.30}$ & $-0.59^{+0.21}_{-0.21}$ & $+0.22^{+0.17}_{-0.17}$ & $+0.50^{+0.20}_{-0.20}$ \\[2mm]
1004135 & $0^{\mathrm{h}}59^{\mathrm{m}}30.62^{\mathrm{s}}$ & $-33^\circ42{}^\prime06.75{}^{\prime\prime}$ & bscl1 & 4669 & 1.51 & -2.09$\pm$0.11 & 0.09 & -0.64 & $+0.31^{+0.13}_{-0.13}$ &  & $<-0.34$ & $-0.57^{+0.27}_{-0.29}$ &  \\[2mm]
1004271 & $0^{\mathrm{h}}59^{\mathrm{m}}31.53^{\mathrm{s}}$ & $-33^\circ39{}^\prime41.73{}^{\prime\prime}$ & bscl1 & 4992 & 2.30 & -1.96$\pm$0.16 & 0.11 & -0.30 &  & $<+2.40$ &  &  &  \\[2mm]
1004320 & $0^{\mathrm{h}}59^{\mathrm{m}}31.95^{\mathrm{s}}$ & $-33^\circ38{}^\prime36.88{}^{\prime\prime}$ & bscl1 & 4874 & 1.80 & -2.04$\pm$0.12 & -0.01 & -2.40 &  & $<+1.11$ &  & $-1.35^{+0.38}_{-0.35}$ & $<+1.45$
\enddata
\tablecomments{Stellar parameters and abundances for the stars in this work. The upper limits given are 2 $\sigma$ upper limits. Here we show the first 10 lines. A full version of this table is available in a machine-readable format.}
\end{deluxetable}
\end{rotatetable}

\begin{figure*}[p]
    \centering
    \includegraphics[width=1\textwidth]{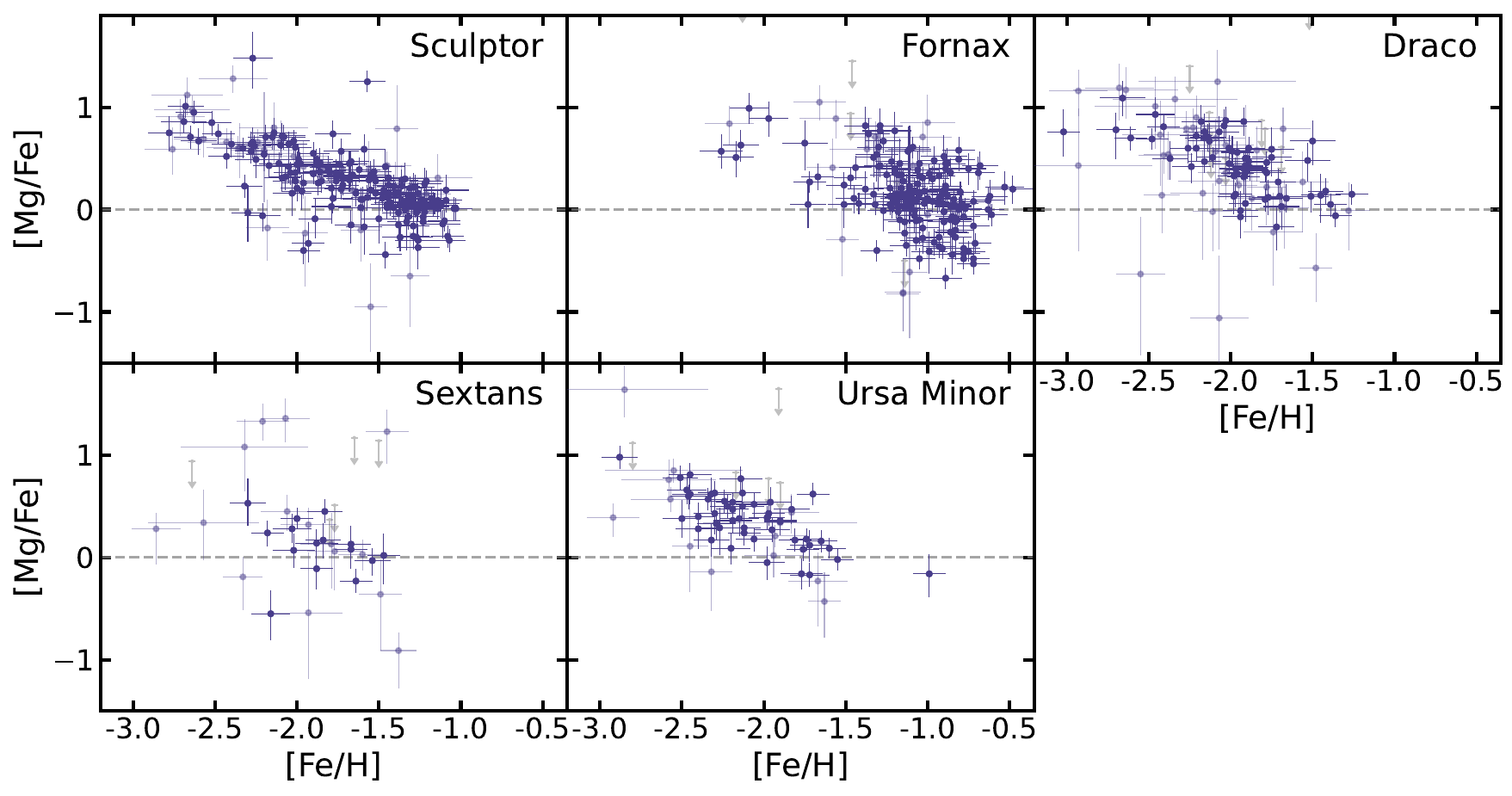}
    \caption{[Mg/Fe] for all five dSphs in our sample. Measurements with upper or lower asymmetric error $>$ 0.3, including systematic error derived in Section \ref{subsec:syserror}, are less opaque. The 2 $\sigma$ upper limits are shown in gray. The dashed line at [Mg/Fe] = 0 is for reference.}
    \label{fig:allMgFe}
\end{figure*}

\begin{figure*}[p]
    \centering
    \includegraphics[width=1\textwidth]{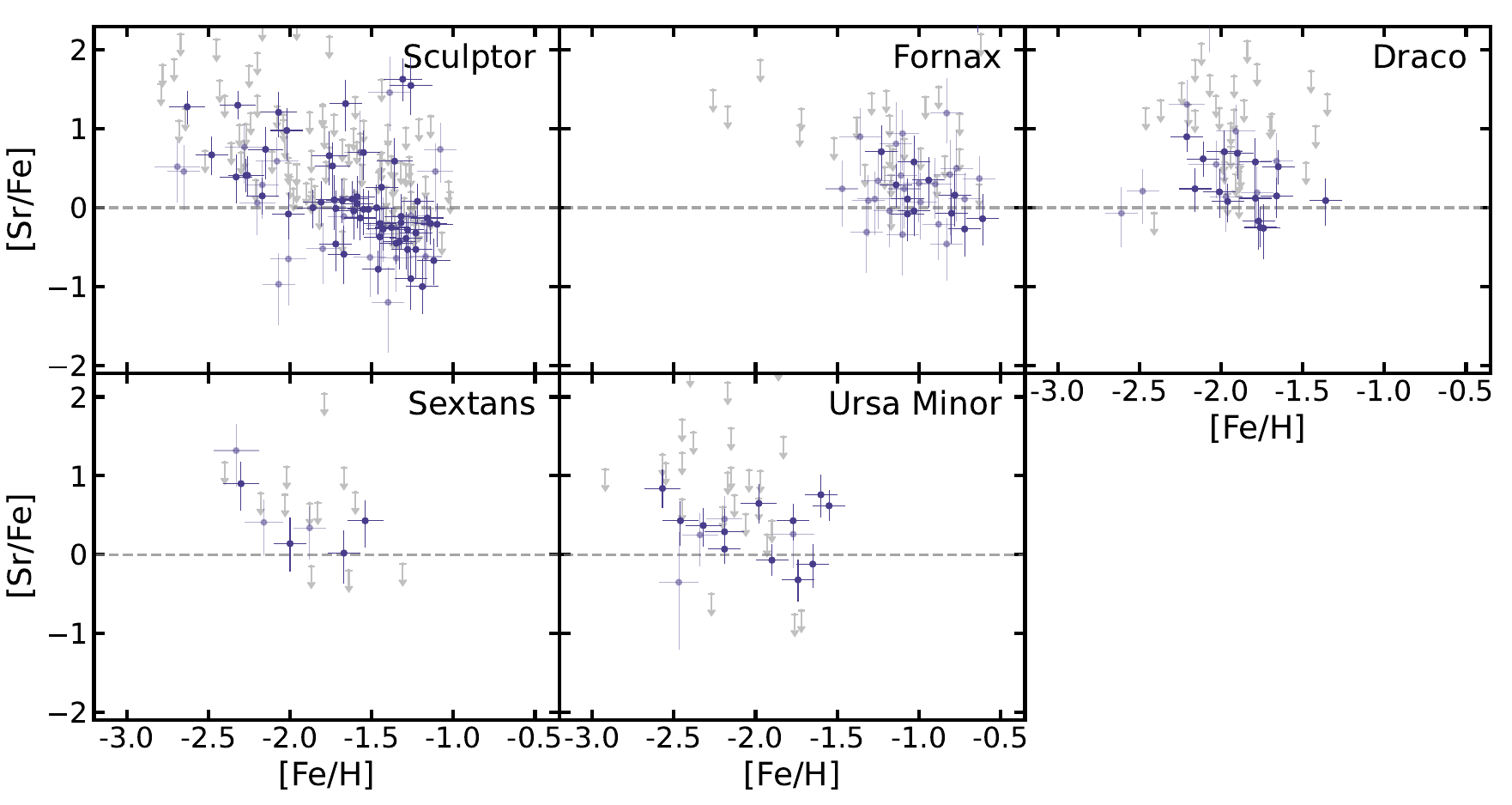}
    \caption{[Sr/Fe] for all five dSphs in our sample. Measurements with upper or lower asymmetric error $>$ 0.4, including systematic error derived in Section \ref{subsec:syserror}, are less opaque. The 2 $\sigma$ upper limits are shown in gray. The dashed line at [Sr/Fe] = 0 is for reference.}
    \label{fig:allSrFe}
\end{figure*}

\begin{figure*}[p]
    \centering
    \includegraphics[width=1\textwidth]{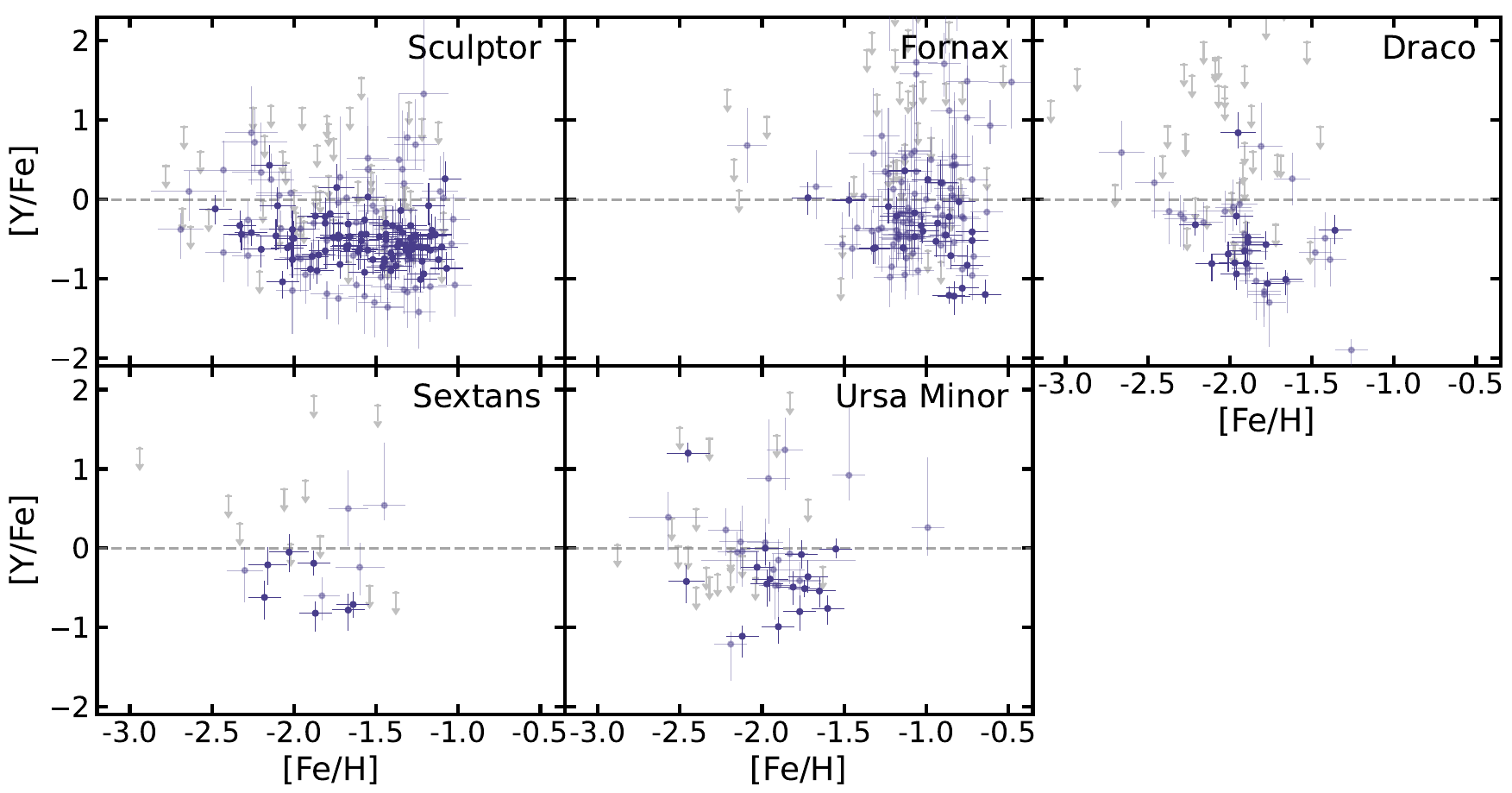}
    \caption{[Y/Fe] for all five dSphs in our sample. Measurements with upper or lower asymmetric error $>$ 0.3, including systematic error derived in Section \ref{subsec:syserror}, are less opaque. The 2 $\sigma$ upper limits are shown in gray. The dashed line at [Y/Fe] = 0 is for reference.}
    \label{fig:allYFe}
\end{figure*}

\begin{figure*}[p]
    \centering
    \includegraphics[width=1\textwidth]{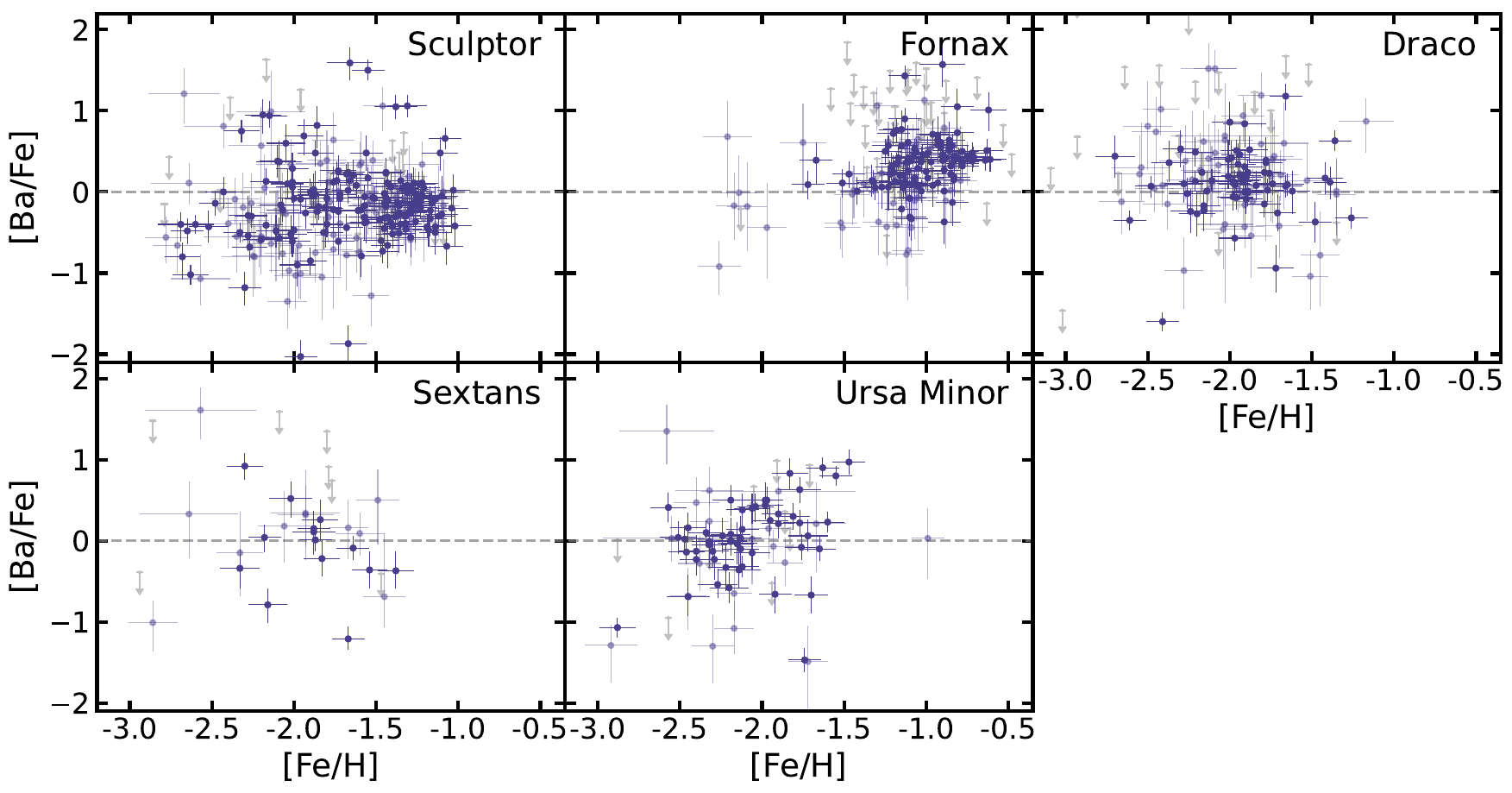}
    \caption{[Ba/Fe] for all five dSphs in our sample. Measurements with upper or lower asymmetric error $>$ 0.3, including systematic error derived in Section \ref{subsec:syserror}, are less opaque. The 2 $\sigma$ upper limits are shown in gray. The dashed line at [Ba/Fe] = 0 is for reference.}
    \label{fig:allBaFe}
\end{figure*}


\begin{figure*}[t]
    \centering
    \includegraphics[width=1\textwidth]{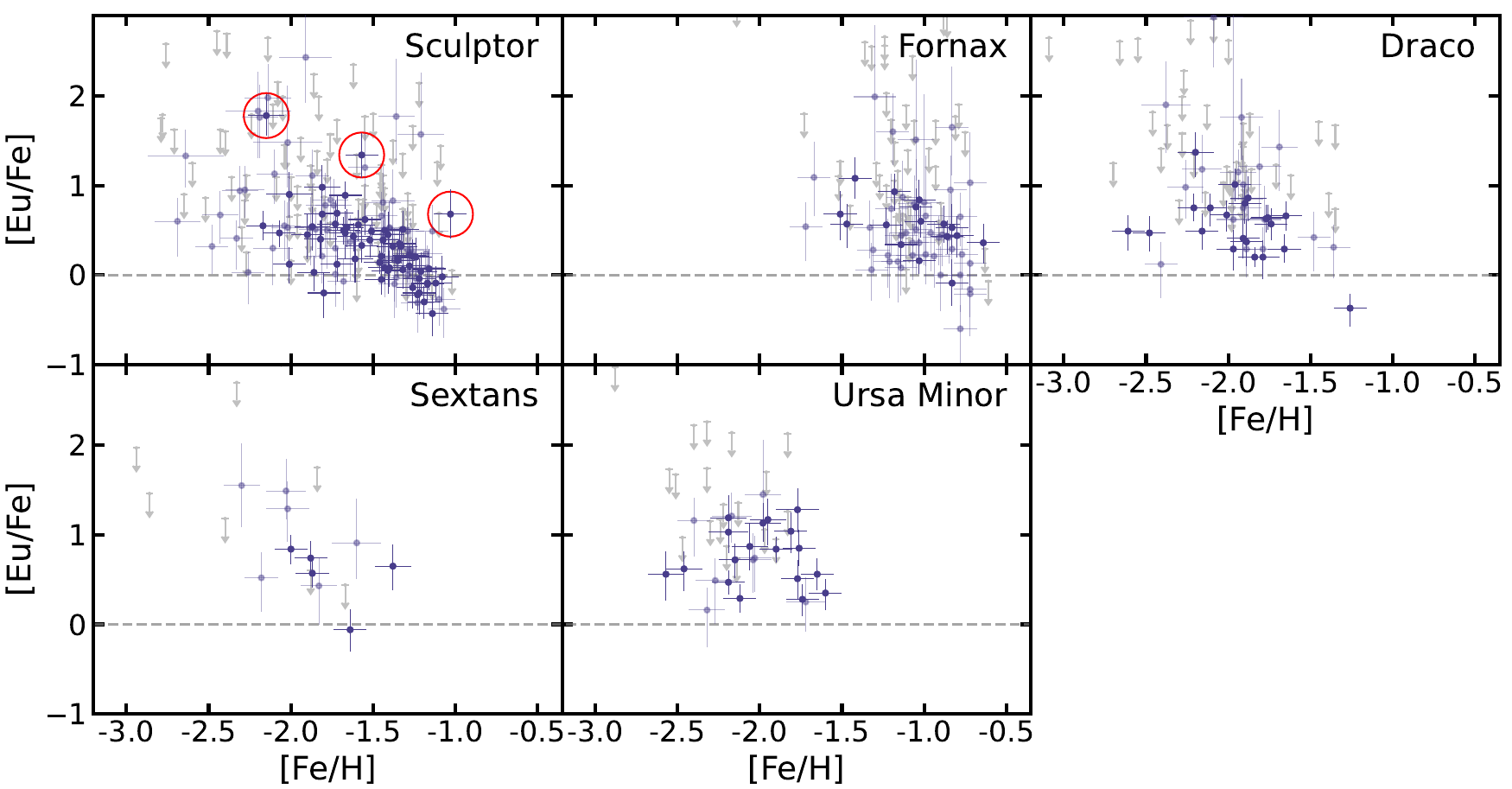}
    \caption{[Eu/Fe] for all five dSphs in our sample. Measurements with upper or lower asymmetric error $>$ 0.3, including systematic error derived in Section \ref{subsec:syserror}, are less opaque. The 2 $\sigma$ upper limits are shown in gray. The dashed line at [Eu/Fe] = 0 is for reference. The Eu-enhanced stars 1013881, 1004838, and 1011213 in Sculptor are circled in red.}
    \label{fig:allEuFe}
\end{figure*}

\section{Results} \label{sec:results} 
Here we present our abundance results for Mg, Sr, Y, Ba, and Eu in the five dSph galaxies in this study, Sculptor, Fornax, Draco, Sextans, and Ursa Minor, and remark on qualitative trends.  Table~\ref{tab:abundances} presents the measured abundances. Figures \ref{fig:allMgFe} - \ref{fig:allEuMg} follow a color scheme to easily differentiate the contents: the dark purple plots show [X/Fe] vs. [Fe/H], the teal plots show [X/Eu] vs. [Fe/H], and the magenta plots show [X/Mg] vs. [Mg/H].

\subsection{The $\alpha$ elements: Mg} \label{subsec:Mg}
Mg abundances for 580 stars in the dSphs are plotted in Figure \ref{fig:allMgFe}. These measurements will provide critical information regarding the relative rates of CCSNe and Type Ia SNe in chemical evolution models. For now, they appear to follow expected patterns.

First, some of the dSphs do not show clear signs of [$\alpha$/Fe]-[Fe/H] knees. The [$\alpha$/Fe]-[Fe/H] knee is the point in a plot of [Mg/Fe] versus [Fe/H] when the abundance pattern transitions from a slope of 0 to a decreasing slope. This is often taken to represent the point when Type Ia SNe start dominating Fe enrichment \citep[e.g.,][]{hendricks_fornax_2014,hill_vltflames_2019}. As pointed out in \citet{tolstoy_star-formation_2009} and further investigated in \citet{mason_realistic_2024}, the shape of the [Mg/Fe] distribution does not only depend on the SN Ia delay time, but on the specific star formation history of the galaxy as well. This means that not all galaxies have obvious [$\alpha$/Fe]-[Fe/H] knees. In fact, \citet{mason_realistic_2024} found that lower mass galaxies, such as dSphs, are more likely than high mass galaxies to have a steadily decreasing [Mg/Fe] instead of a knee.

The other notable feature in Figure \ref{fig:allMgFe} is the variety of spreads in [Mg/Fe] at any given metallicity in the dSphs. For example, at [Fe/H] = -1.5, Fornax has $\approx$1 dex spread in [Mg/Fe] while Sculptor has a narrower distribution with a spread of $\approx$0.5 dex. This indicates that Fornax had a longer and more complex star formation history than other dSphs \citep{kirby_multi-element_2011-1, weisz_star_2014}.

\subsection{First-peak s-process: Sr and Y} \label{subsec:light-s}
In this study, the light $s$-process elements (and first-peak $r$-process elements) are represented by Sr and Y\@. At low metallicities, the production of these elements should be dominated by the $r$-process, and possibly $s$-process contributions from fast-rotating massive stars \citep[FRMSs,][]{pignatari_s-process_2008, chiappini_FRMS_2011, frischknecht_non-standard_2012, choplin_non-standard_2018}, as there has not been enough time for the $\sim3~M_{\sun}$ AGB stars that make most of the $s$-process elements \citep[e.g.,][]{karakas_slar_2016} to evolve. At higher metallicities, Sr and Y are dominated by $s$-process production.

Figures \ref{fig:allSrFe} and \ref{fig:allYFe} show the [Sr/Fe] and [Y/Fe] values for the stars in this study. Sculptor appears to have higher [Sr/Fe] values at low metallicity. For stars with low [Fe/H], [Sr/Fe] must be high to be detectable with our abundance measurement technique, which might make [Sr/Fe] appear higher than it actually is at low metallicity. This potential bias emphasizes the importance of including the upper limit measurements in Figures \ref{fig:allSrFe} and \ref{fig:allYFe}.


\subsection{Barium}\label{subsec:heavy-s}
Like Sr and Y, Ba production is dominated by the $s$-process at late times but is likely primarily produced by prompt $r$-process events at low metallicities. Our [Ba/Fe] measurements largely match those of \citet{duggan_neutron_2018}, which is expected because some of the measurements are from the same observations and we are using the same set of Ba lines.

Sculptor has a large scatter in [Ba/Fe] in Sculptor at [Fe/H] $<$ -1.5, possibly because the early $r$-process sources are rare and therefore stochastic (Figure \ref{fig:allBaFe}). The scatter decreases at higher metallicities when AGB stars become the primary source of Ba. 

 In Fornax [Ba/Fe] increases over the metallicity range. The weighted Pearson correlation coefficient for [Ba/Fe] v.\ [Fe/H] in Fornax is 0.22$\pm$0.05, with the inverted variances of [Ba/Fe] constituting the weights because that is the dominant error \citep{rodgers_pearsonr_1988}. \footnote{The weighted Pearson correlation coefficient values and errors given here and elsewhere in Sections \ref{sec:results} and \ref{sec:discussion} are estimated from Monte Carlo bootstrapping with the error on the $y$-axis determining the weight. Largely independent measurement errors in abundances are expected to partially decorrelate the observed relationships. For this reason, the Pearson correlation coefficient is best understood as a lower bound for how correlated the abundances are.} This positive slope can be explained if Fornax had a longer, more gradual SFH than the other dSphs. The abundances in Fornax would reflect a large contribution from AGB stars because more stars would be formed after AGB stars start producing $s$-process elements. This is discussed further in Section \ref{subsec:s-process} regarding Figure \ref{fig:allBaMg}.


\subsection{Europium} \label{subsec:Eu}
Europium is almost entirely produced in the $r$-process \citep[this is known to be true for the Sun in particular;][]{simmerer_rise_2004,sneden_neutron-capture_2008, kobayashi_origin_2020}. This makes it a very good indicator of $r$-process delay times and enrichment in dSphs. Sculptor, Fornax, Draco, and Sextans all have higher [Eu/Fe] at low metallicity (Figure \ref{fig:allEuFe}). As mentioned above, Fe production becomes significant with the onset of Type Ia SNe. At this point, [Eu/Fe] starts to decrease rapidly as Eu production does not increase with Fe production. As with the $s$-process elements, [Eu/Fe] also has a large scatter at low metallicities. 

\begin{figure*}[t]
    \centering
    \includegraphics[width=1\textwidth]{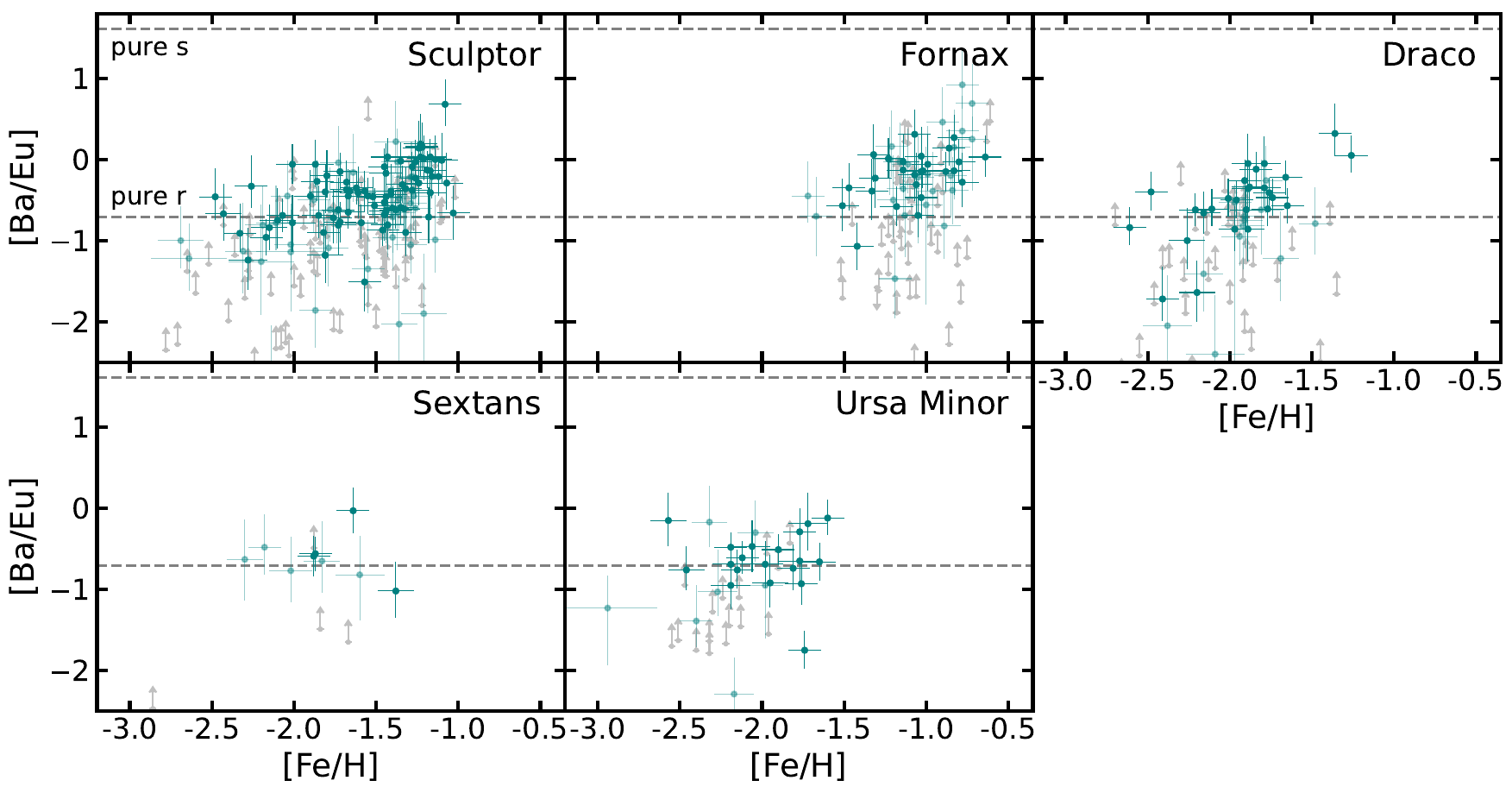}
    \caption{[Ba/Eu] for all five dSphs in our sample. Measurements with Ba and Eu combined upper or lower asymmetric error $>$ 0.4, including systematic error derived in Section \ref{subsec:syserror}, are less opaque. The 2 $\sigma$ upper limits for [Eu/Fe] or [Ba/Fe] are plotted in gray. The horizontal dashed lines represent the [Ba/Eu] expected for pure $s$-process (${\rm [Ba/Eu]} = +1.611$) and $r$-process (${\rm [Ba/Eu]} = -0.708$) enrichment. These values were calculated using the $s$- and $r$-process abundances from \citet{simmerer_rise_2004} and the solar abundances from \citet{asplund_chemical_2009}.}
    \label{fig:allr-s}
\end{figure*}

\begin{figure*}[t]
    \centering
    \includegraphics[width=1\textwidth]{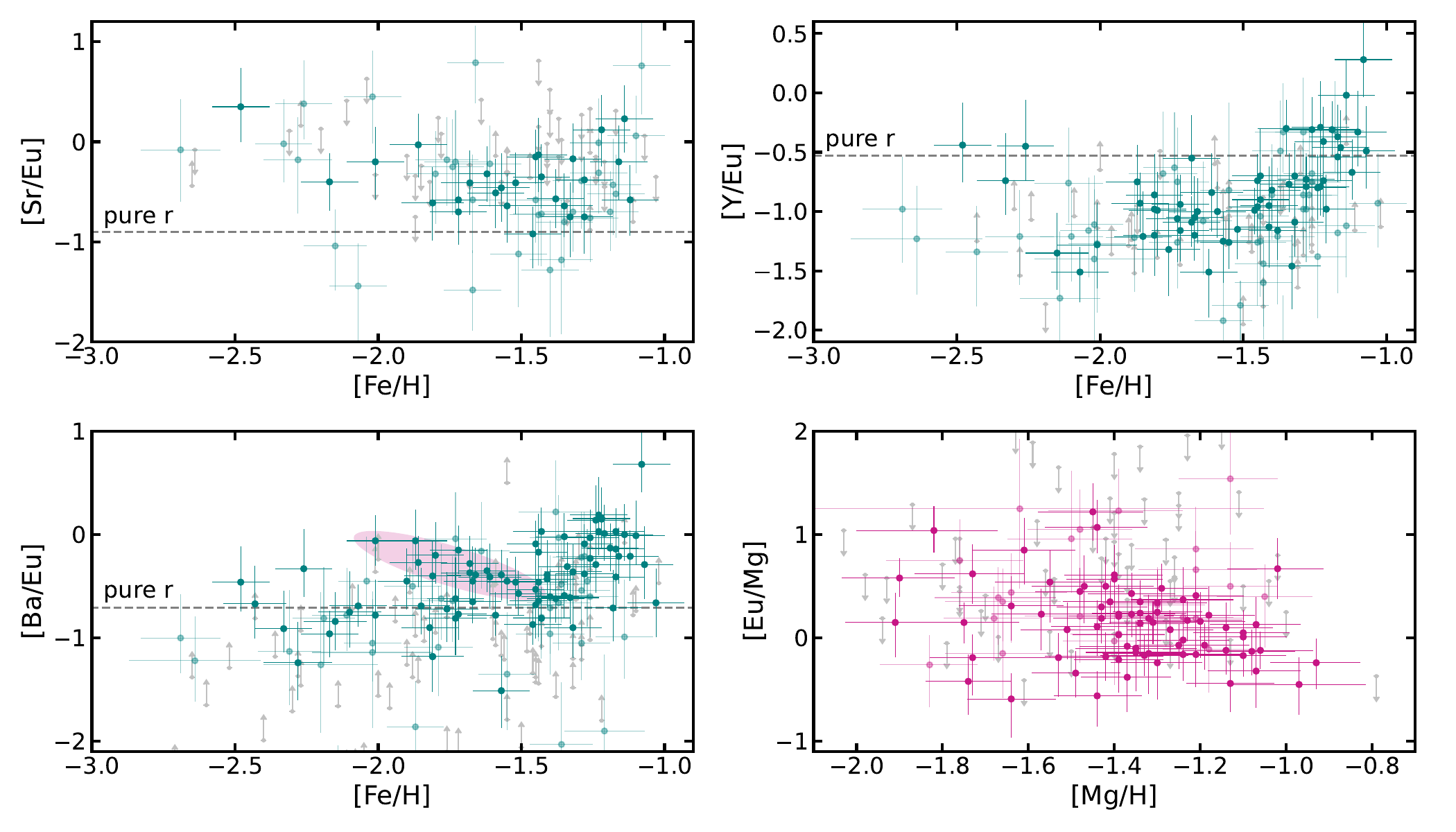}
    \caption{[Sr/Eu], [Y/Eu], and [Ba/Eu] vs.\ [Fe/H] and [Eu/Mg] vs.\ [Mg/H] for Sculptor. Measurements with upper or lower asymmetric error combined for both elements used in the plot $>$ 0.4, including systematic error derived in Section \ref{subsec:syserror}, are less opaque. The 2 $\sigma$ upper limits are plotted in gray. Stars with upper limits in both element abundances in the panel are not included. The horizontal dashed lines represent the [Sr/Eu], [Y/Eu], or [Ba/Eu] expected for pure $r$-process enrichment (${\rm [Sr/Eu]} = -0.897$, ${\rm [Y/Eu]} = -0.527$, and ${\rm [Ba/Eu]} = -0.708$). These values were calculated using the $s$- and $r$-process abundances from \citet{simmerer_rise_2004} and the solar abundances from \citet{asplund_chemical_2009}. The pink ellipse in the lower left panel highlights the metal-poor stars inconsistent with pure $r$-process enrichment discussed in Section \ref{subsec:s-r}.}
    \label{fig:SclEu}
\end{figure*}

\begin{figure}
    \centering    \includegraphics[width=1\linewidth]{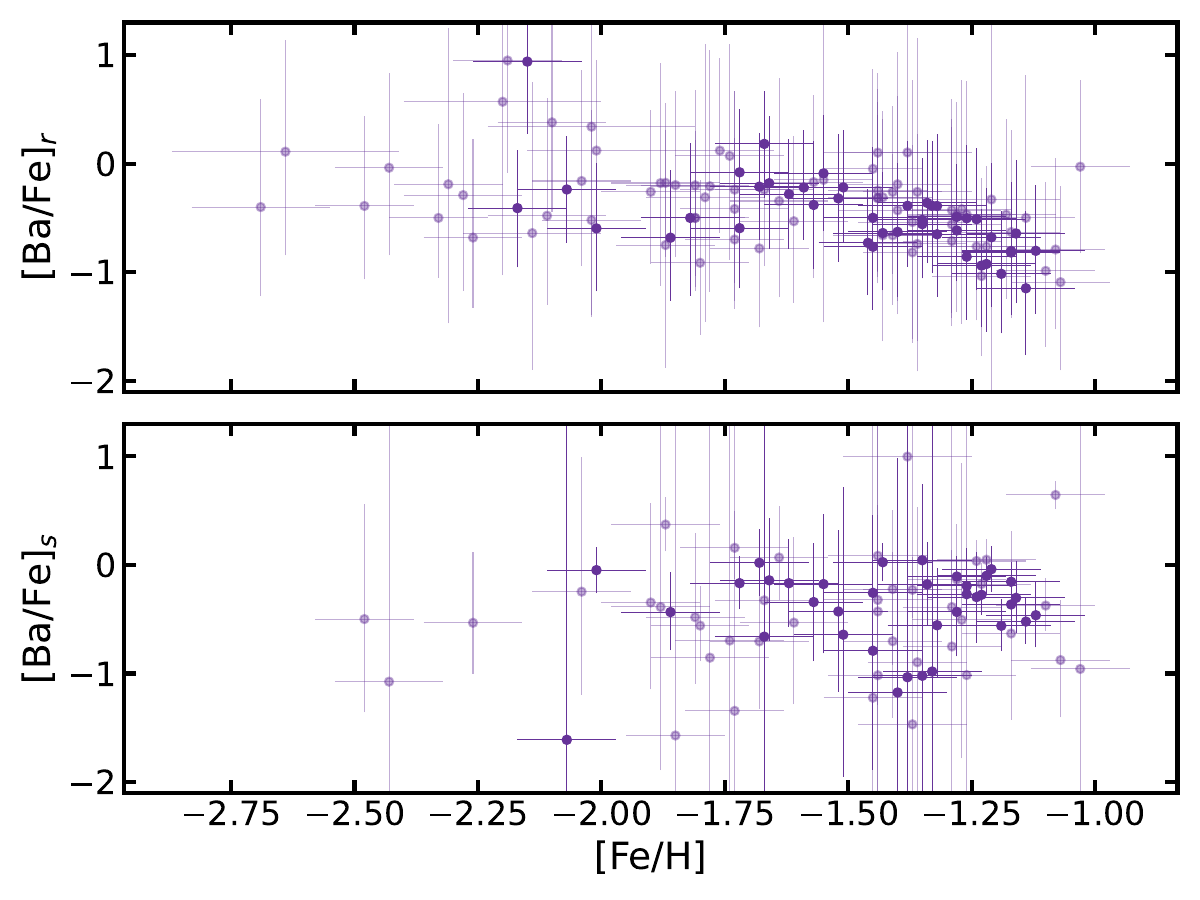}
    \caption{$r$- (top panel) and $s$-process (bottom panel) [Ba/Fe] in Sculptor. The $f_r$ and $f_s$ fractions are calculated using Equation~\ref{eq:fr}. Plotted error is calculated using standard propagation of error. This is an overestimate because the error on Ba, including $\delta_{\rm sys}$, appears twice in the equation for the error. Measurements with combined Ba and Eu averaged upper or lower asymmetric error $>$ 0.3 are less opaque. Upper limits are not plotted because they cannot be used to calculate $f_r$ and $f_s$.}
    \label{fig:Bar-s}
\end{figure}

\section{Discussion} \label{sec:discussion} 
Here we expand on the qualitative implications of the abundance trends we observe. These trends reflect the star formation histories and neutron-capture nucleosynthesis in the dSphs in our sample.  A quantitative interpretation should be made with the benefit of chemical evolution models, which we defer to future work.

\subsection{The evolution of the s- and r-process in dSphs} \label{subsec:s-r}
The ratio of a primarily $s$-process element to an $r$-process element, such as [Ba/Eu], can be used to compare the contributions of $s$- and $r$-process neutron-capture nucleosynthesis. Figure \ref{fig:allr-s} shows [Ba/Eu] for the dSphs, with horizontal lines representing the [Ba/Eu] expected for pure $s$- and $r$-process nucleosynthesis as defined by the solar abundances \citep{simmerer_rise_2004, asplund_chemical_2009}. As expected, at low metallicities [Ba/Eu] hovers around the pure $r$-process line, as there has not been enough time for intermediate-mass stars to evolve onto the asymptotic giant branch and start contributing significant amounts of $s$-process material. At higher metallicites, once AGB stars start producing large amounts of Ba, [Ba/Eu] starts to approach the pure $s$-process line.

The dSphs show a diversity of times (or metallicities) for the onset of $s$-process nucleosynthesis. Fornax does not have as many stars that appear around the pure $r$-process line as in Sculptor. Instead, [Ba/Eu] starts increasing almost immediately. This difference could be attributed to differences in their star formation histories. If Sculptor had a large burst of star formation early on, more stars would be formed before AGB stars started contributing significant amounts of $s$-process elements. Likewise, Fornax's more gradual, extended star formation period \citep[e.g.,][]{weisz_star_2014} engendered fewer stars with neutron-capture elements from only the $r$-process and resulted in fewer low-metallicity stars overall \citep{kirby_multi-element_2011}. Draco, on the other hand, does display a transition to increasing [Ba/Eu], but at a lower metallicity than Sculptor: ${\rm [Fe/H]} \approx -2$ in Draco as opposed to $\approx -1.5$ in Sculptor. We interpret these neutron-capture abundance patterns as evidence of a steeper age-metallicity relation in Sculptor than in Draco. When AGB stars began producing Ba, increasing the [Ba/Eu] ratio, Sculptor was already at a higher overall metallicity than Draco.

Since Sculptor has so many stars in our sample and a large metallicity range, it is worth looking closer at its $r$- and $s$-process comparison. To do this, we plot [Sr/Eu], [Y/Eu], and [Ba/Eu] in Figure \ref{fig:SclEu}. As in Figure \ref{fig:allr-s}, we include a horizontal line at the pure $r$-process value. 

In Sculptor, there is an unexpected spread in [Ba/Eu] in the region $-2.0 < {\rm [Fe/H]} < -1.5$. For a given value of [Fe/H], there are stars with low ${\rm [Ba/Eu]} \sim -0.7$ and stars that appear to be enhanced in $s$-process elements up to ${\rm [Ba/Eu]} \sim 0.0$. The stars in this region that are enhanced in $s$-process elements compared to other stars with the same metallicity are highlighted by the pink ellipse in Figure \ref{fig:SclEu}.
These apparently $s$-process-polluted stars do not fit within the simple nucleosynthesis schema that would predict constant [X/Eu] around the pure $r$-process value at low metallicities followed by an increase in [X/Eu] with the onset of AGB $s$-process production. Without comparing these measurements to chemical evolution models, it is difficult to determine the cause of this feature. It is possible that it is a signature of an $s$-process contribution with a shorter delay time than AGBs, such as FRMSs. This feature in [Ba/Eu] is not apparent in the other dSphs. However, there are fewer measurements for the other dSphs, they cover a smaller metallicity range, and they have different SFHs from Sculptor.


Ideally, we would isolate the $s$- and $r$-process components of the neutron-capture abundances to study the sites of nucleosynthesis separately. To this end, we used the Eu abundances in the stars to calculate $r$- and $s$-process fractions following Equation~\ref{eq:fr} for an element X, such as Ba \citep{duggan_neutron_2018}. $f_r({\rm X},{\rm Eu})$ represents the fraction of element X that is produced in the $r$-process, not the fraction of all material in that star produced by the $r$-process.\footnote{Consider Eu.  Nearly all Eu in almost any star is formed primarily by the $r$-process, even if the star is heavily polluted by the $s$-process.}  By corollary, $f_s = 1 - f_r$. 

\begin{equation}
    f_r({\rm X},{\rm Eu}) = \frac{\frac{N_{{\rm Eu}_s}}{N_{{\rm X}_s}} - 10^{{\rm [X/Eu]}+({\rm X/Eu})_{\odot}}}{\frac{N_{{\rm Eu}_s}}{N_{{\rm X}_s}} - \frac{N_{{\rm Eu}_r}}{N_{{\rm X}_r}}} \label{eq:fr}
\end{equation}

\noindent $N_{{\rm X}_s}$ and $N_{{\rm X}_r}$ are the solar system $s$- and $r$-process abundances from \citet{simmerer_rise_2004}. We used this equation to calculate $\rm{[Ba/Fe]}_r$ and  $\rm{[Ba/Fe]}_s$. We focus on Ba in Sculptor because we were able to get reliable Ba abundances for the most stars in Sculptor. Furthermore, Sculptor covers the largest metallicity range of the dSphs in our sample. Figure \ref{fig:Bar-s} shows the results of the $r$- and $s$-process fraction calculations for Ba in Sculptor. The $\rm{[Ba/Fe]}_r$ values trace the same pattern as [Eu/Fe]. $\rm{[Ba/Fe]}_s$ has the most points at higher metallicities because at low metallicities, $f_s$ is essentially zero. The significant increase in stars with $\rm{[Ba/Fe]}_s$ measurements between ${\rm [Fe/H]} \approx -1.75$ and $\approx -1.50$ marks the point when AGB $s$-process nucleosynthesis becomes the primary source of Ba in Sculptor.

\begin{figure*}[p]
    \centering
    \includegraphics[width=1\textwidth]{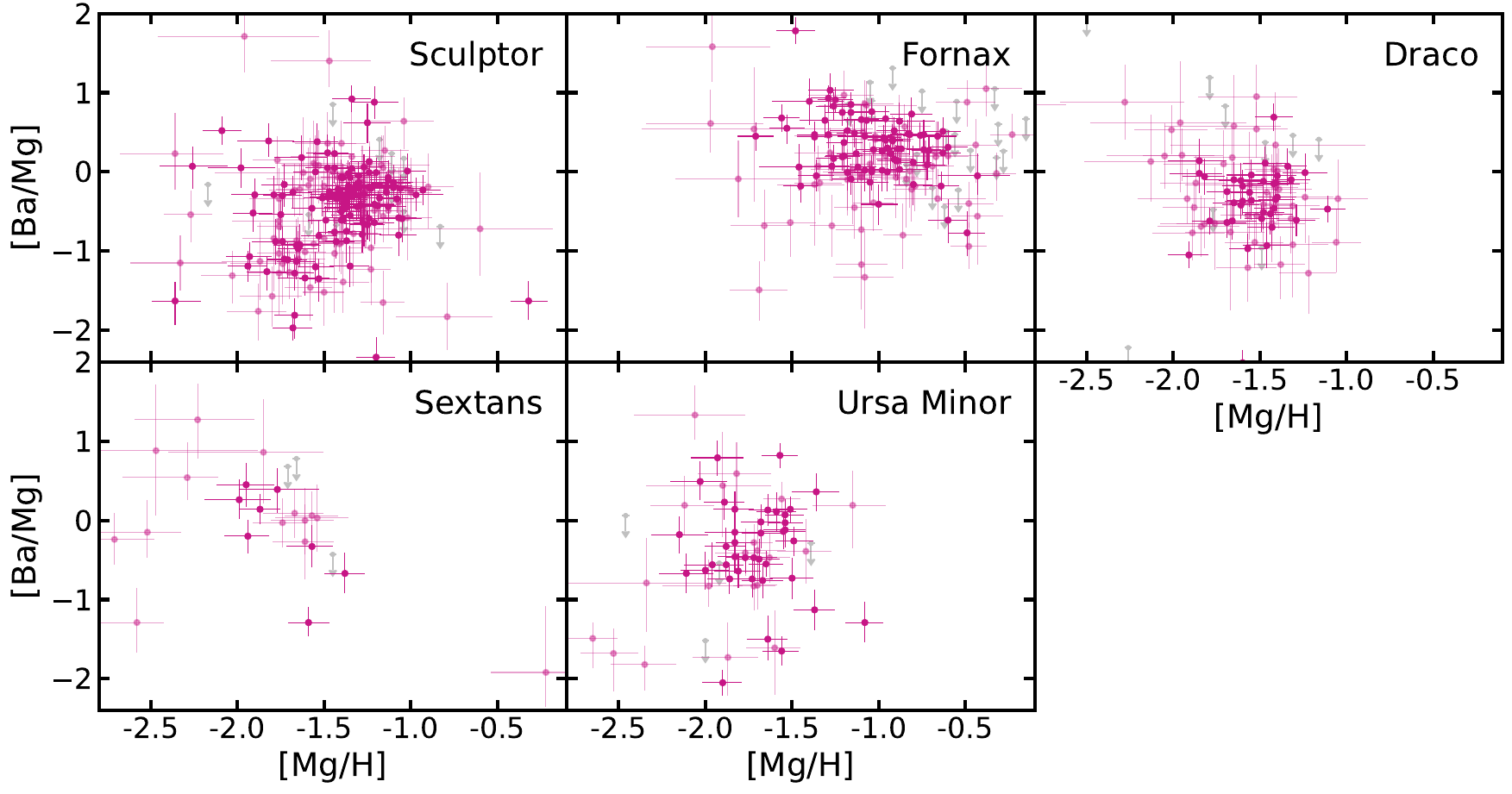}
    \caption{[Ba/Mg] for all five dSphs in our sample. Measurements with Ba and Mg combined upper or lower asymmetric error $>$ 0.3, including systematic error derived in Section \ref{subsec:syserror}, are less opaque. The 2 $\sigma$ upper limits for [Ba/Fe] are shown plotted in gray.}
    \label{fig:allBaMg}
\end{figure*}

\subsection{Implications for the s-process} \label{subsec:s-process}
The $s$-process elements in our study span a large range of atomic number, from the first-peak (Sr, Y) to the lanthanides (Ba). This gives us the opportunity to compare different facets of the $s$-process in multiple dSphs. 

While Figure \ref{fig:Bar-s} displays the onset of AGB $s$-process contributions (see Section \ref{subsec:s-r}), it features some $\rm{[Ba/Fe]}_s$ at low metallicity in Sculptor. This may be additional evidence for an $s$-process production site that operates on a short timescale, such as FRMSs. However, there are many other factors that can impact the $\rm{[Ba/Fe]}_s$ we observe, such as the assumption of solar system neutron-capture abundances from \citet{simmerer_rise_2004} that might not describe dSphs perfectly.

Another way to investigate the $s$-process in these galaxies is to plot [Ba/Mg] as a function of [Mg/H] (Figure \ref{fig:allBaMg}). As observed by \citet{skuladottir_neutron-capture_2019}, [Ba/Mg] increases in Sculptor with increasing [Mg/H]\@. The weighted Pearson correlation coefficient for this relationship is 0.15$\pm$0.04. At first glance, this trend is expected, especially at high metallicity. Ba is produced by AGB stars, which have a longer delay time than the CCSNe that produce Mg. This relationship produces a positive slope in [Ba/Mg].

Interestingly, Fornax and Draco do not reflect this trend. Sextans appears to have a negative slope, but has a much smaller sample size than the other dSphs. These differences in [Ba/Mg] are further evidence that the abundance distributions in these galaxies depend on both the delay-times of various species and their star formation histories, and that dSphs do not have identical evolutionary histories.





\begin{figure*}[p]
    \centering
    \includegraphics[width=1\textwidth]{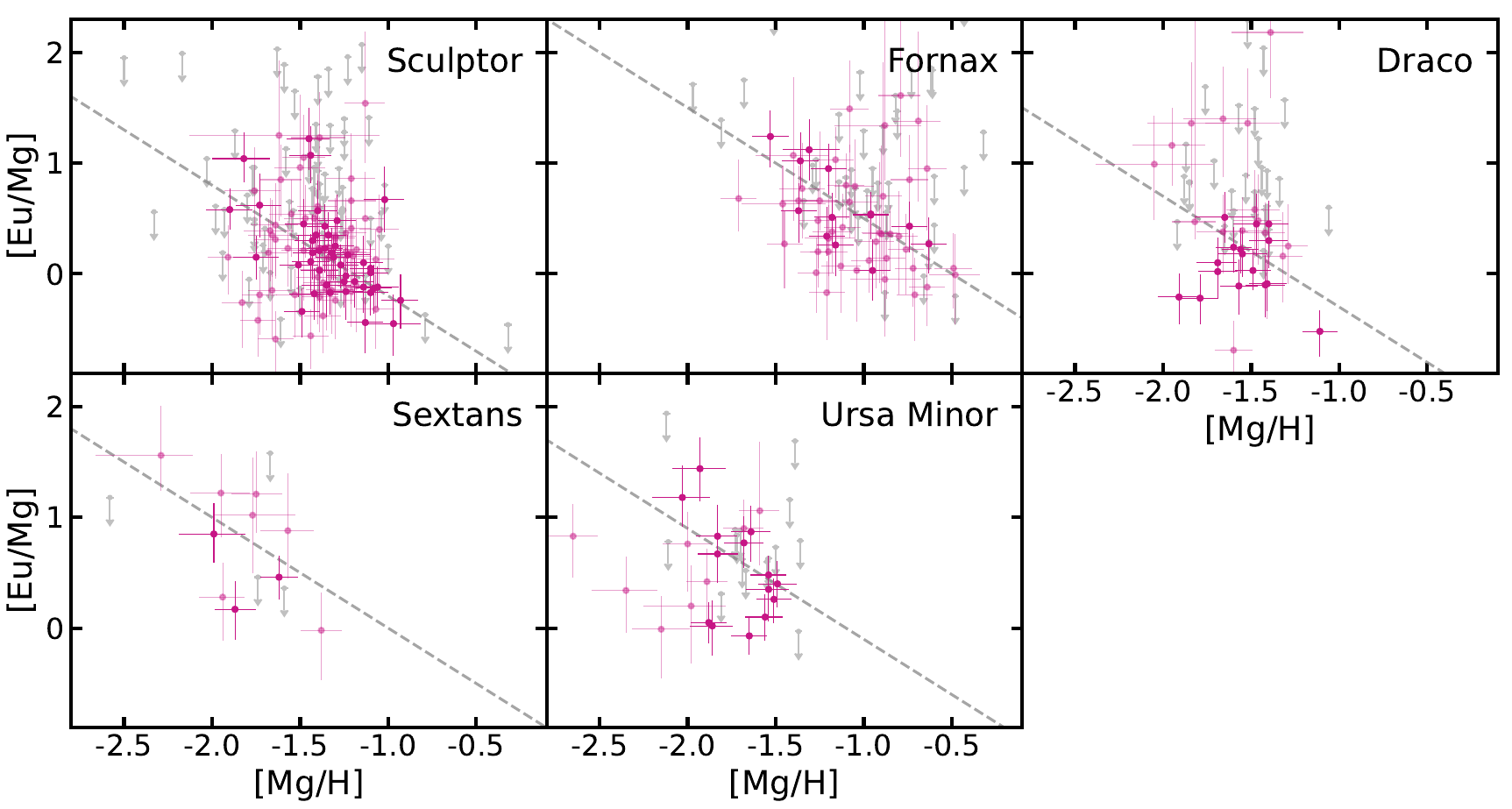}
    \caption{[Eu/Mg] for all five dSphs in our sample. Measurements with Eu and Mg combined upper or lower asymmetric error $>$ 0.3, including systematic error derived in Section \ref{subsec:syserror}, are less opaque. The 2 $\sigma$ upper limits for [Eu/Fe] are shown plotted in gray. The gray dashed line is $y = -x$ shifted down to intercept the data and is meant to serve as a point of comparison of the slope in the abundances of [Eu/Mg].}
    \label{fig:allEuMg}
\end{figure*}

\subsection{Implications for the r-process} \label{subsec:r-process}
Since we are able to measure [Eu/Fe] from medium-resolution spectra, we can learn about the $r$-process from the large sample of Eu abundances in this study. In particular, we explore here the implications for the timescale on which the $r$-process elements are produced.

Figure \ref{fig:allEuMg} compares the Eu abundances to Mg in the dSphs; Figure \ref{fig:SclEu} features a closer look at this relationship for Sculptor in the bottom right panel. The trend of [Eu/Mg] as a function of [Mg/H] can be used to investigate how the timescale of Eu production compares to the fast delay time of CCSNe. Exotic types of CCSNe are thought to be possible sites of the $r$-process \citep{argast_r-process_2004,nishimura_r-process_2015}. If the slope of [Eu/Mg] vs.\ [Mg/H] is flat, then Eu and Mg have similar delay times, and some amount of Eu must be produced by $r$-process events with a timescale similar to CCSNe.

All five dSphs show decreasing [Eu/Mg] as a function of [Mg/H]. The weighted Pearson correlation coefficient for [Eu/Mg] v.\ [Mg/H] is $-0.21 \pm 0.07$ in Sculptor and $-0.30 \pm 0.09$ in Fornax, the two dSphs for which we have the most measurements. This anti-correlation suggests that Mg is delayed with respect to Eu and that at least some Eu is made in processes that have a very short delay time, such as special types of SNe. Our results corroborate those of \citet{skuladottir_neutron-capture_2019} and \citet{molero_evolution_2021}, who concluded that neutron star mergers are likely not the dominant source of the $r$-process in dSph galaxies, at least at low metallicities.

There is an alternative explanation for the markedly negative slopes in [Eu/Mg] vs.\ [Mg/H] seen in Fornax, Sextans, and Ursa Minor. It could be that almost all of the Eu production in those galaxies occurred very quickly after star formation started and then ended abruptly. CCSNe continued to produce Mg, which increased [Mg/H] at the same rate as it decreased [Eu/Mg].  This would produce a slope of $-1$ between [Eu/Mg] and [Mg/H], which is plotted in the background of Figure \ref{fig:allEuMg}.  Although the scatter in [Eu/Mg] is large, the slopes for all five dSphs is roughly consistent with $-1$.  Therefore, we conclude that the majority of Eu production in the galaxies occurred at early times, corresponding to ${\rm [Mg/H]} \sim -2$.

Disentangling the possible explanations of the trends seen here in [Eu/Mg] will require detailed comparison with galactic chemical evolution models, which is beyond the scope of this paper. For now, we simply conclude that $r$-process enrichment in dSphs requires a nucleosynthesis site in addition to neutron star mergers and that this site has a short delay time.

\subsection{Outliers} \label{subsec:outliers}
The large sample size and metallicity range of the stars in this study is useful not only for identifying trends in the populations, but also in identifying interesting outliers. Though medium-resolution spectroscopy is not best-suited for measuring absolute abundances, it can be used to find stars that are good candidates for high-resolution follow-up.

Three such stars in Sculptor are circled in Figure \ref{fig:allEuFe}. Stars 1013881, 1004838, and 1011213, have higher [Eu/Fe] than their counterparts at similar metallicities: [Eu/Fe] = $1.78$, $1.34$, and $0.68$, respectively. At low metallicity, it is not particularly surprising to find a star enhanced in $r$-process elements. Since $r$-process events are expected to be rare and stochastic, some stars may end up being more enriched than others. Similarly at late times, as the star formation in the galaxy is using the last of the gas, the ejecta from an $r$-process event would not have much gas to mix with and could thus nearly single-handedly enrich a star with a large concentration of $r$-process material \citep{mcwilliam_evidence_2018,atzberger_leiptr_2024}. A Eu-enhanced star at high metallicity, possibly enriched in this way, has been observed in Sculptor by \citet{geisler_sculptor-ing_2005}. The presence of a Eu-enhanced star at moderate metallicity (${\rm [Fe/H]} = -1.57$ for 1004838) is more surprising. This star is also intriguing because contrary to what is expected for an $r$-process enriched star, it does not appear to be enriched in Ba; it has [Ba/Eu] = $-1.51^{+0.33}_{-0.36}$, well below the pure $r$-process line at ${\rm [Ba/Eu]} = -0.708$ in Figures \ref{fig:allr-s} and \ref{fig:SclEu}.

The most metal-poor star of the three already has high-resolution measurements from VLT/UVES, analyzed by \citet{reichert_neutron-capture_2020}. They found that this star is enhanced in Eu as well. The other two outliers circled in Figure~\ref{fig:allEuFe} do not already have archival high-resolution spectra available. It would be interesting to perform high-resolution follow-up on these two stars to confirm whether they are indeed enhanced in Eu and whether their abundances of other neutron-capture elements are remarkable. Unfortunately, they are not very bright, with V magnitudes of $\approx$18.7.

Some stars show an enhancement in $s$-process abundances, most easily seen in [Ba/Fe] in Figure~\ref{fig:allBaFe}. As noted by \citet{duggan_neutron_2018}, these stars were likely polluted by an AGB binary companion at some point in their evolution. In fact, two such stars in Sculptor, 1013218 and 1016224, have [Ba/Eu] values consistent with $s$-process pollution: [Ba/Eu] = $0.50^{+0.60}_{-0.37}$ and $0.22^{+0.50}_{-0.38}$, respectively. Enrichment from a binary companion would cause an increase in $s$-process abundances that is unrelated to the chemical evolution of the galaxy.





\section{Conclusion} \label{sec:conclusion}
The primary conclusions of this paper are as follows:
\begin{enumerate}
    \item Neutron-capture abundances measured from medium-resolution spectra agree within the errors with the same abundances measured from higher resolution spectra. This opens the door to measuring neutron-capture elements in many more stars than is possible with high-resolution spectroscopy alone.
    \item The [Eu/Fe] and [Eu/Mg] abundances in our sample (Figures \ref{fig:allEuFe}, \ref{fig:SclEu}, and \ref{fig:allEuMg}) suggest that at least some Eu is produced on timescales similar to or faster than the typical CCSNe that produce the majority of Mg. Thus, it is unlikely that neutron star mergers are the sole source of $r$-process enrichment in dSphs.
    \item $\rm{[Ba/Fe]}_s$, [Sr/Eu], [Y/Eu], and [Ba/Eu] in Sculptor (Figures \ref{fig:SclEu} and \ref{fig:Bar-s}) show evidence for $s$-process elements produced with a delay time shorter than that of AGB stars. Perhaps this is a signature of the $s$-process in FRMSs.
\end{enumerate}

We are able to observe these trends because of the large sample size and metallicity range made possible by using medium-resolution spectra to measure neutron-capture abundances.  Our work bodes well for massively multiplexed spectrographs, such as Subaru/PFS \citep{takada_pfs_2014}, 4MOST \citep{dejong_4most_2019}, and VLT/MOONS \citep{cirasuolo_moons_2020}.

There are many opportunities for future work based on this study. For example, it would be very interesting to apply this method to study other dwarf spheroidal galaxies, such as Carina, or to increase the limited sample size in ultra-faint dwarf galaxies.

As we pointed out in Section \ref{sec:discussion}, quantitative comparison with galactic chemical evolution models is imperative to draw conclusions about the nucleosynthesis in these dSphs based on these observations \citep[e.g.][]{marcolini_chemical_2008,kirby_multi-element_2011-1,homma_chemical_2015,molero_evolution_2021,delosreyes_simultaneous_2022}.  In a subsequent paper, we will use the precisely-known delay time of AGB stars and the plentiful $s$-process abundances in this work to calculate star formation histories of these dSphs. The high resolution of these SFHs, thanks to the use of $s$-process abundances, might be able to clarify the extent to which reionization in the early universe stopped or slowed star formation in small galaxies.





\vspace{2em}
Some of the data presented herein were obtained at Keck Observatory, which is a private 501(c)3 non-profit organization operated as a scientific partnership among the California Institute of Technology, the University of California, and the National Aeronautics and Space Administration. The Observatory was made possible by the generous financial support of the W. M. Keck Foundation.  The authors wish to recognize and acknowledge the very significant cultural role and reverence that the summit of Maunakea has always had within the Native Hawaiian community. We are most fortunate to have the opportunity to conduct observations from this mountain.


%

\vspace{5mm}

\facility{Keck:II (DEIMOS)}


\software{\texttt{astropy} \citep{astropy_I, astropy_II, astropy_III},
\texttt{ATLAS9} \citep{kurucz_new_1993},
\texttt{linemake} \citep{placco_linemake_2021, placco_linemake_2021-1},
\texttt{matplotlib} \citep{Hunter_matplotlib_2007},
\texttt{MOOG} \citep{sneden_MOOG_2012},
\texttt{scipy} \citep{scipy}, 
\texttt{spec2d} pipeline \citep{cooper_spec2d_2012,newman_deep2_2013}
    }


\bibliography{SMAUG_refs_ADS}{}
\bibliographystyle{aasjournal}



\end{document}